\documentclass[]{pasj01}

\Received{2020/05/06}
\Accepted{2020/08/08}
 
\usepackage{graphicx} 
\usepackage{natbib} 
\begin{document} 

\title{ 
Precision radial velocity measurements by the forward-modeling technique in the near-infrared
\thanks{Based on data collected at Subaru Telescope, which is operated by the National Astronomical Observatory of Japan.}}

\author{Teruyuki \textsc{hirano}\altaffilmark{1}}
\altaffiltext{1}{Department of Earth and Planetary Sciences, Tokyo Institute of Technology,
2-12-1 Ookayama, Meguro-ku, Tokyo 152-8551, Japan}
\email{hirano@geo.titech.ac.jp}

\author{Masayuki \textsc{kuzuhara}\altaffilmark{2,3}}
\altaffiltext{2}{Astrobiology Center, NINS, 2-21-1 Osawa, Mitaka, Tokyo 181-8588, Japan}
\altaffiltext{3}{National Astronomical Observatory of Japan, NINS, 2-21-1 Osawa, Mitaka, Tokyo 181-8588, Japan}

\author{Takayuki \textsc{kotani}\altaffilmark{2,3,4}}
\altaffiltext{4}{Department of Astronomy, School of Science, The Graduate University for Advanced Studies (SOKENDAI), 2-21-1 Osawa, Mitaka, Tokyo, Japan}

\author{Masashi \textsc{omiya}\altaffilmark{2,3}}

\author{Tomoyuki \textsc{kudo}\altaffilmark{5}}
\altaffiltext{5}{Subaru Telescope, 650 N. Aohoku Place, Hilo, HI 96720, USA}
\author{Hiroki \textsc{harakawa}\altaffilmark{5}}
\author{S\'ebastien \textsc{vievard}\altaffilmark{2,5}}

\author{Takashi \textsc{kurokawa}\altaffilmark{2,6}}
\altaffiltext{6}{Institute of Engineering, Tokyo University of Agriculture and Technology, 2-24-16, Nakacho, Koganei, Tokyo, 184-8588, Japan}

\author{Jun \textsc{nishikawa}\altaffilmark{2,3}}

\author{Motohide \textsc{tamura}\altaffilmark{2,3,7}}
\altaffiltext{7}{Department of Astronomy, Graduate School of Science, The University of Tokyo, 7-3-1 Hongo, Bunkyo-ku, Tokyo 113-0033, Japan}

\author{Klaus \textsc{hodapp}\altaffilmark{8}}
\altaffiltext{8}{University of Hawaii, Institute for Astronomy, 640 N. Aohoku Place, Hilo, HI 96720, USA}

\author{Masato \textsc{ishizuka}\altaffilmark{7}}

\author{Shane \textsc{jacobson}\altaffilmark{8}}

\author{Mihoko \textsc{konishi}\altaffilmark{9}}
\altaffiltext{9}{Faculty of Science and Technology, Oita University, 700 Dannoharu, Oita 870-1192, Japan}

\author{Takuma \textsc{serizawa}\altaffilmark{6}}

\author{Akitoshi \textsc{ueda}\altaffilmark{3}}

\author{Eric \textsc{gaidos}\altaffilmark{10}}
\altaffiltext{10}{Department of Earth Sciences, University of Hawaii at M\={a}noa, Honolulu, HI 96822, USA}

\author{Bun'ei \textsc{sato}\altaffilmark{1}}




\maketitle

\begin{abstract}
Precision radial velocity (RV) measurements in the near infrared are a powerful tool 
to detect and characterize exoplanets around low-mass stars or young stars with higher
magnetic activity. However, the presence of strong telluric absorption lines 
and emission lines 
in the near infrared that significantly vary in time can prevent extraction of RV
information from these spectra by classical techniques, which ignore or mask 
the telluric lines. 
We present a methodology and pipeline to derive precision RVs from near-infrared spectra 
using a forward-modeling technique. We applied this to spectra 
with a wide wavelength coverage ($Y$, $J$, and $H$ bands, simultaneously), 
taken by the InfraRed Doppler (IRD) spectrograph on 
the Subaru 8.2-m telescope. Our pipeline extracts the instantaneous instrumental 
profile of the spectrograph for each spectral segment, based on a reference spectrum 
of the laser-frequency comb that is injected into the spectrograph simultaneously with 
the stellar light. 
These profiles are used to derive the intrinsic stellar template spectrum,
which is free from instrumental broadening and telluric features, as well as model 
and fit individual observed spectra in the RV analysis. 
Implementing a series of numerical simulations using theoretical spectra 
that mimic IRD data, we test the pipeline and show that IRD can achieve 
$<2$ m s$^{-1}$ precision for slowly rotating mid-to-late M dwarfs with a signal-to-noise 
ratio $\gtrsim 100$ per pixel at 1000 nm. Dependences of RV precision 
on various stellar parameters (e.g., $T_\mathrm{eff},\,v\sin i,\,[\mathrm{Fe/H}]$) and the 
impact of telluric-line blendings on the RV accuracy are discussed through the mock 
spectra analyses. 
We also apply the RV-analysis pipeline to the observed spectra of GJ 699 and TRAPPIST-1, 
demonstrating that the spectrograph and the pipeline are capable of an RV accuracy of $<3$ 
m s$^{-1}$ at least on a time scale of a few months. 
\end{abstract}

\section{Introduction}
M dwarfs are drawing increasing attention in exoplanet studies, for their ubiquity and advantages to search for small planets. 
The combination of low effective temperatures and small radii of M dwarfs makes the 
habitable zone (HZ) for exoplanets much closer to their host stars \citep[e.g.,][]{2016ApJ...819...84K},
which facilitates the detection and characterization of small planets in the HZ. 
Recent transit surveys from the space ({\it Kepler, K2}) have unveiled the population of 
small planets around M dwarfs in unprecedented detail 
\citep[e.g.,][]{2013ApJ...767...95D, 2015ApJ...807...45D, 2016ApJ...816...66B, 2016MNRAS.457.2877G, 2017AJ....154..207D}, 
whose properties are similar, but sometimes distinct from those of planets around solar-type stars
\citep[e.g.,][]{2013ApJ...767...95D, 2018AJ....155..127H}.

Most previous planet searches around M dwarfs by Doppler observations have been 
conducted with high-resolution optical spectrographs 
\citep[e.g.,][]{2013A&A...549A.109B, 2017A&A...602A..88A}. 
To take advantage of the characteristic of M dwarfs that they are brighter in the near infrared
(NIR), new types of Doppler observations have recently been attempted with NIR
high-resolution spectrographs, such as CARMENES at 
Calar Alto 3.5-m telescope \citep{2016SPIE.9908E..12Q}, the Habitable-Zone Planet Finder
\citep[HPF:][]{2014SPIE.9147E..1GM} on the Hobby Eberly Telescope, 
and SPIRou on the CFHT 3.58-m telescope \citep{2014SPIE.9147E..15A}. 
Some of those NIR instruments are shown to achieve good radial velocity (RV) precisions 
for M dwarfs similar to those by optical spectrographs\footnote{See e.g., https://hpf.psu.edu}.

With a goal of finding small planets in or near the HZ around M dwarfs and characterizing 
those planets in terms of mass, orbit, and atmosphere, we developed a NIR 
high-resolution instrument, the InfraRed Doppler (IRD) spectrograph, which was installed 
on the Subaru 8.2-m telescope in 2017 
\citep{2012SPIE.8446E..1TT, 2014SPIE.9147E..14K, 2018SPIE10702E..11K}. 
IRD is a fiber-fed, stabilized spectrograph, which can simultaneously cover from $930$ 
up to $1740$ nm ($Y$, $J$, and $H$ bands) with a spectral resolution of $R\approx 70,000$. 
Stellar light collected by the telescope is first focused by the adaptive optics 
system \citep[AO188:][]{2008SPIE.7015E..10H} and injected into the first fiber through the 
fiber injection module. 
A second fiber can be connected to the spectrograph, into which we usually inject the comparison 
(wavelength calibration) light from the laser-frequency comb 
\citep[LFC:][]{2016OExpr..24.8120K, 2016SPIE.9912E..1RK}. 
The LFC spectrum consists of a large number of emission lines whose positions are separated 
by a fixed interval in the frequency domain. The LFC spectrum simultaneously injected with 
the stellar light is used to correct for any instrumental wavelength (velocity) drift and
variations of the point spread function of the spectrograph, which is indispensable for 
precision and accurate RV measurements by high-resolution spectroscopy.

In this paper, we present our technique and algorithm to derive precision RVs from the 
NIR spectra, especially acquired by Subaru/IRD. 
RV measurement techniques have been studied and developed by many groups for a number 
of spectrographs, most of which exploit cross-correlation based techniques 
\citep[e.g.,][]{2002A&A...388..632P, 2013A&A...549A.109B} or forward-modeling ones with 
the least-squares fitting 
\citep[e.g.,][]{1996PASP..108..500B, 2002PASJ...54..873S, 2010ApJ...713..410B, 2012ApJS..200...15A, 2018A&A...609A..12Z}. 
Unlike visible spectra ($\lambda\lesssim 700$ nm), however, NIR spectra are heavily 
contaminated by telluric absorption and emission lines, which can vary in shape due to temporal variations of atmospheric and sky conditions. 
These variations of telluric lines can lead to large systematic errors in the derived RVs 
unless taken into account in RV measurements. 
In the {\tt SERVAL} pipeline \citep{2018A&A...609A..12Z}, developed for optical and NIR 
spectra taken e.g., by CARMENES \citep{2016SPIE.9908E..12Q}, telluric lines positions are
completely masked in the RV fitting procedure. 
Yet, given that there are very limited ranges of the spectrum (e.g.,
$990\,\mathrm{nm}<\lambda<1070\,\mathrm{nm}$) 
that are nearly free from telluric absorptions in the whole wavelength region covered 
by IRD ($970\,\mathrm{nm}<\lambda<1720\,\mathrm{nm}$), 
this masking scheme for all the telluric lines does not work for the RV analyses of 
IRD spectra. This difficulty in handling the telluric lines in the NIR, as well as the instrumental characteristics of IRD (e.g., the use of the LFC for a simultaneous reference), made us develop our own pipeline 
to derive RVs from NIR spectra by the forward-modeling technique in the presence of 
time-variable telluric lines.

The rest of this paper is organized as below. 
In Section \ref{s:obs}, we briefly describe the observations of several standard stars using
Subaru/IRD and present the data reduction to extract one-dimensional (1D) spectra. 
Section \ref{s:method} presents the detailed descriptions on our scheme and methodology to derive
precision RVs from the IRD spectra. 
We first test the RV pipeline using theoretical (mock) spectra which are generated 
based on the properties of the IRD spectrograph and theoretically synthesized M-dwarf spectra (Section \ref{s:mock}). 
In the same section, we discuss the RV precisions achievable for different types of stars 
(e.g., spectral type, rotation velocity, stellar metallicity, etc) through a series of Monte 
Carlo simulations. To demonstrate the on-sky performances of IRD, 
we analyze the data of GJ 699 and TRAPPIST-1 using our pipeline, and Section \ref{s:onsky}
summarizes the results. 
Section \ref{s:discussion} is devoted to the summary and discussion, in which we also 
present the future prospects to improve the pipeline.

\section{Observations and Data Reduction}\label{s:obs}
We carried out observations of several standard stars (RV and telluric standards) with Subaru/IRD 
during IRD's engineering nights in 2018 and open-use programs in 2018 and 2019
to test and demonstrate IRD's on-sky performances. 
Targets of observations are explained in more detail in Section \ref{s:onsky}.
For each observing run (typically a few to 10 nights), we took flat-lamp frames for each of the two fibers. 
For the comparison spectra, we used both Thorium-Argon (Th-Ar) hollow cathode lamp and LFC.

\begin{figure*}
\centering
\includegraphics[width=15cm]{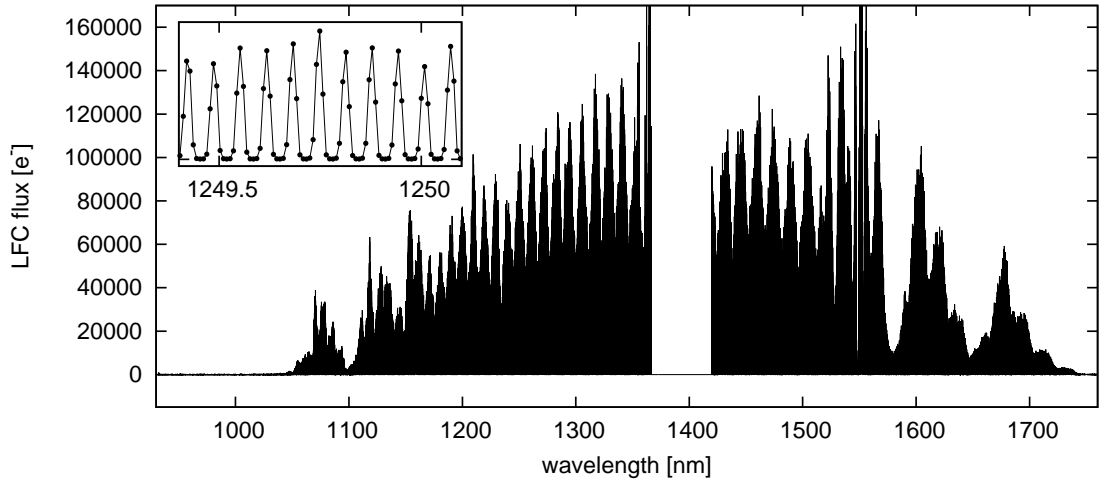}
\caption{
A sample of LFC spectrum taken by Subaru/IRD with the integration time of 300 sec. 
Each mountain-shaped feature (periodic ) corresponds and reflects the blaze function of
each echelle order. 
The inset displays a small region of the same spectrum ($1249.4\,\mathrm{nm}<\lambda<1250.1\,\mathrm{nm}$). 
}
\label{fig:comb}
\end{figure*}
Raw IRD frames were reduced by the standard procedure for echelle data reductions. 
Since the HAWAII-2RG detectors used in IRD are known to show count biases that are dependent 
on each readout channel, 
we used our custom code to suppress the bias counts \citep{2018SPIE10702E..60K}. 
The subsequent reduction steps (flat fielding, scattered light subtraction, extraction 
of one-dimensional spectra, and wavelength calibration with the Th-Ar lamp) were performed 
using the echelle package of \texttt{IRAF}. 
Approximate wavelength calibration was done based on emission lines of the Th-Ar 
comparison spectrum \citep{Kerber_2008_ThAr}, which covers the whole range of IRD spectra. 
For precise RV measurements, however, both the precision and accuracy of the wavelength calibration are not good enough to achieve $\approx 1$ m s$^{-1}$ due to the lack of strong Th-Ar lines in the NIR, and thus we recalibrated the wavelengths using the LFC spectrum 
for each fiber. For the recalibration of wavelength, we fitted individual emission lines 
of LFC by Gaussians and identified peak positions for all available lines. 
We then reassigned 
the wavelength to each pixel so that the peak positions are exactly separated by $12.5$ GHz 
in the frequency domain (= designed separation of LFC emission lines
$\approx 0.042$ nm at 1000 nm). 
LFC spectrum currently covers between $1050$ nm and $1720$ nm (Figure \ref{fig:comb}), thus
we were capable of this recalibration for only those wavelengths. Below 1050 nm, 
we kept the wavelength solutions determined by the Th-Ar lamp and we did not use those 
shorter wavelengths for RV measurements in Section \ref{s:onsky}.

\section{Methodology}\label{s:method}

In this section, we briefly introduce the concept of RV measurements using simultaneous 
observations of a reference spectrum and illustrate the characteristics of the 
IRD spectrograph, with which we reach the conclusion that
we need forward modeling to derive accurate RVs from IRD spectra. 
Our RV analysis pipeline is described in detail in subsection \ref{s:pipeline}. 

\subsection{Simultaneous Reference Technique}\label{s:simul}

In the simultaneous reference technique, 
precision stellar RVs are usually measured in two steps. 
First, using the wavelength-calibrated 1D stellar 
spectrum, absolute RVs from the positions of absorption lines are measured. 
This is often done by cross-correlating the observed spectrum against a
template spectrum; the most successful optical spectrographs 
(e.g., ESO/HARPS) adopt box-shaped numerical masks
as template spectra \citep[e.g.,][]{2002A&A...388..632P}. 
Let us denote this measured RV by $v_\mathrm{obs}$. 
The second step is the measurement of the ``drift" of the spectrum due to environmental
variations of the spectrograph (i.e., temperature and pressure). 
The spectral drift of order $10^{-3}$ pixel usually translates to an apparent RV 
variation of $1-2$ m s$^{-1}$; 
the correction of this small pixel drift is essential to achieve extreme precisions. 
In order to assess the spectral drift, 
the wavelength-reference spectrum (e.g., from a Th-Ar lamp) is obtained simultaneously 
and cross-correlated against a template wavelength-reference spectrum. 
Under the assumption that the pixel (or velocity) drift of the stellar spectrum is 
exactly the same as that of the reference spectrum, the RV drift $v_\mathrm{drf}$ is estimated 
by fitting the cross-correlation function (CCF) by e.g., a single Gaussian and measuring 
the center of the CC function. 
The final stellar RV $v_\star$ is then determined by
\begin{eqnarray}
\label{eq:1}
v_\star = v_{\rm obs} - v_{\rm drf}. 
\end{eqnarray}
For IRD, 
the LFC, injected into the reference fiber, is designed to correct for the instrumental drifts
by precise tracking of the LFC line positions. 
See Figure \ref{fig:comb} for a sample LFC spectrum taken with a 5-minute integration.

Since it is essential to set the offset between the wavelength solutions for both fibers 
to zero, for the absolute wavelength calibration we split the calibration-source 
(either Th-Ar lamp or LFC) light 
into the two fibers and inject those into the IRD spectrograph simultaneously;
if those calibration sources are not observed simultaneously for the two fibers, 
the temporal drift of spectrum positions leads to a systematic relative offset 
between the two wavelength solutions for the two fibers. 
This offsetting needs to be done for each IRD observing run, as the position of the IRD 
echellogram is also known to move along the ``spatial" direction on the detectors 
over a time scale of a few months to a year, due to variations in environmental conditions 
(Kotani et al. in prep.).   
Every time a new set of calibration data with the LFC (or Th-Ar) being injected into both fibers
is obtained, wavelengths are calibrated based on the procedure described in 
Section \ref{s:obs} for each fiber.

\subsection{Instrumental RV Drift of IRD}\label{s:IRD}

\begin{figure}
\includegraphics[width=8.5cm]{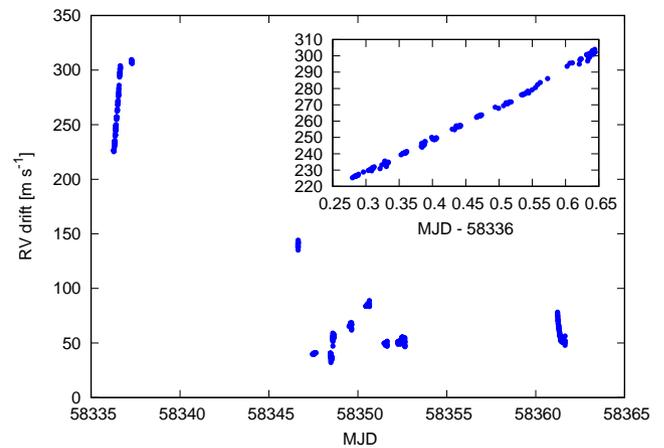}
\caption{
RV drift (representative value for all spectral segments) of the IRD spectrograph during a month. 
The RV variation within a night is plotted in the inset. 
}
\label{fig:abs}
\end{figure}
\begin{figure}
\includegraphics[width=8.5cm]{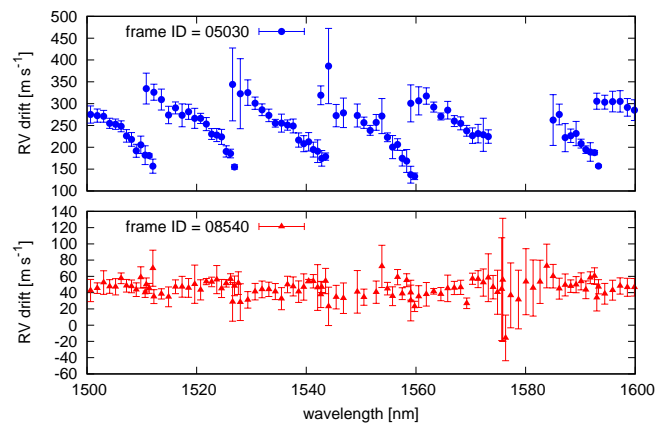}
\caption{
RV drifts measured at individual spectral segments (each $\Delta \lambda = 0.7- 1$ nm) for two different frames taken in 2018 August. 
}
\label{fig:abs2}
\end{figure}
IRD exhibits temperature variations, typically a few 10 mK to over 100 mK (peak-to-valley), depending on the position of the measurement point. 
This temperature variation, probably arising from electrical drift of the temperature 
sensors and/or temperature controller, was found to cause a large apparent RV drift in time. 
Figure \ref{fig:abs} plots the observed RV drift
of IRD, measured using the LFC spectra taken during 2018 August; We computed the RV drift 
of each LFC spectrum by cross-correlating it against a template LFC spectrum, which is 
generated by combining a large number ($>50$) of LFC frames taken during a relatively 
short time interval. As shown in the figure, the RV drift could be as large as $\approx 300$ m s$^{-1}$,
and RVs can vary by $> 50$ m s$^{-1}$ even during a single night. 
Similar RV drifts measured from LFC data were also reported in \citet{2018SPIE10702E..60K}.
We found that the RV drift of the spectrograph is strongly correlated with temperature 
variations of the camera lens just in front of the detectors (Kotani et al. in prep.).

In addition to this overall RV drift (i.e., averaged over all orders) in time, 
the apparent RV drift of the LFC spectrum depends on the position on the detector; when we split the whole spectrum into many small spectral segments
(e.g., $\approx 1$ nm), each spectral segment was found to have a different RV drift value, which is more evident when the overall (mean) RV drift is large. 
Figure \ref{fig:abs2} displays the results of segment-by-segment drift measurements 
for two different IRD frames. In this figure, we split each echelle order into 19 segments 
and fitted each segment with LFC emission lines
to the template LFC spectrum by the least-squares technique. 
The positional dependence of the RV drift is more significant when the overall RV drift 
is large (ID = 05030; upper panel), while the RV drift takes similar values for all 
segments (different orders) when the overall RV is relatively small 
(ID = 08540; lower panel). 
The primary reason for this positional dependence is that each pixel on the IRD detectors 
does not cover the same interval in wavelength nor velocity, and the parallel pixel shift 
of the spectrum results in a different degree of drift in the velocity. For example, 
for the echelle order of 95, one pixel covers $\Delta \lambda\approx 0.0055$ nm at the
shortest-wavelength edge of the order, but it corresponds to 
$\Delta \lambda\approx 0.0135$ nm at the other edge of the detector. This implies that 
a pixel shift of e.g., $0.1$ pixel corresponds to the velocity drifts of 
$\approx 110$ m s$^{-1}$ at the shortest-wavelength edge and $\approx 260$ m s$^{-1}$ 
at the longest-wavelength edge of the same order, respectively. This fact qualitatively 
explains the behaviors in Figure \ref{fig:abs2} at least as an 
approximation\footnote{More precise effects of the spectrum shift are under investigation.}. 
The fact that the wavelength (velocity) coverage is different for each pixel on the detector 
suggests that a temperature instability of the spectrograph leads not only 
to a parallel shift of the spectrum against wavelength or velocity, but also to an 
effective width (shape) variation of the ``instrumental profile" (IP) in the velocity domain 
for each spectral segment; therefore Equation (\ref{eq:1}) cannot be used to correct for 
the instrumental drift for the case of IRD. 

\subsection{RV Analysis Pipeline of IRD}\label{s:pipeline}

\begin{figure*}
\centering
\includegraphics[width=12cm]{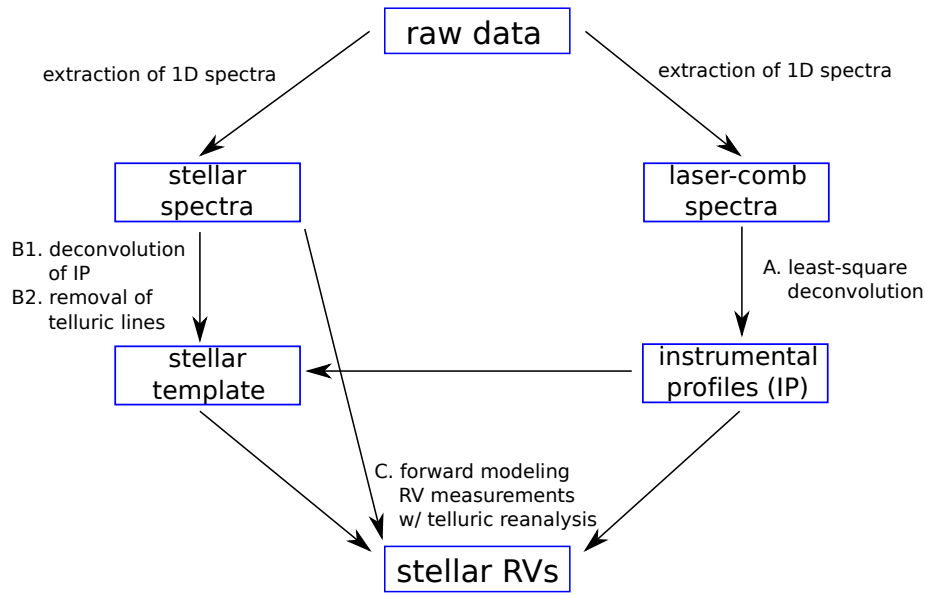}
\caption{
Flowchart of the RV analysis pipeline for Subaru/IRD. 
}
\label{fig:chart}
\end{figure*}



Besides the instrumental challenges above, strong telluric spectral features are 
a major issue for NIR RV measurements. The telluric imprints can significantly vary 
depending on the target's airmass and observing conditions (humidity, in particular). 
In NIR spectroscopy, standard stars have been traditionally observed immediately before 
and/or after the scientific exposures to correct for the telluric lines, but it is unrealistic 
in a Doppler survey to point to standard stars for every exposure of a target.

Based on all these technical challenges and characteristics of the IRD instrument, 
we decided to 
adopt a forward-modeling technique in the RV analysis for IRD data. 
RV measurements at visible wavelengths using forward-modeling techniques have been discussed 
in the literature, especially for measurements with the iodine absorption cell
\citep[e.g.,][]{1996PASP..108..500B, 2002PASJ...54..873S}. 
Figure \ref{fig:chart} shows the flow chart of the IRD data reduction and RV measurements. 
Instead of employing the classical CCF technique often used in the optical 
simultaneous-reference technique, we attempt to correct for the impacts 
of time-variable telluric lines and IPs by modeling an observed NIR spectrum $f_\mathrm{obs}(\lambda)$ as
\begin{eqnarray}
\label{eq:2}
f_\mathrm{obs} (\lambda) = k(\lambda) \nonumber\\
\times\left[S\left(\lambda\sqrt{\frac{1+v_\star/c}{1-v_\star/c}}\right)T\left(\mathbf{A};\lambda\sqrt{\frac{1+v_{\rm tel}/c}{1-v_{\rm tel}/c}}\right) \right] 
*\mathrm{IP},~~
\end{eqnarray}
where $*$ represents the convolution operator, and
$S(\lambda)$ is the intrinsic stellar spectrum (free of telluric lines), Doppler-shifted 
by the stellar RV $v_\star$. The telluric absorption spectrum $T(\mathbf{A}; \lambda)$ includes 
a few relevant telluric parameters $\mathbf{A}$ 
(i.e., precipitable water vapor amount, target airmass). 
The telluric velocity-shift $v_{\rm tel}$ is introduced so as to take into account 
a possible small variation of telluric line positions, due to e.g., winds.
The factor $k(\lambda)$ represents the overall normalization of the spectral continuum, 
which we express by a quadratic function of wavelength $\lambda$. 
IP denotes the instrumental profile of the spectrograph, whose shape depends on wavelength 
(= position on the detector).

One advantage of the forward-modeling RV measurement by Equation (\ref{eq:2}) is that
since the exact wavelength positions of all LFC lines are known {\it a priori}, 
any relative instrumental RV drifts as well as variations in the spectrograph's
point-spread function can be precisely 
traced by extracting the instantaneous IP from the LFC spectrum; 
the extracted IP for each spectral segment should contain those pieces of
information, and the overall velocity shift of the extracted IP's centroid corresponds to the instrumental RV drift of that segment.
Hence, by convolving this extracted IP in modeling an observed stellar spectrum
(Equation (\ref{eq:2})), one no longer needs to subtract segment-by-segment 
instrumental drifts as in Equation (\ref{eq:1}). In this methodology, 
it is important to use ``fixed" wavelength solutions for all frames\footnote{Note that wavelength solutions are different for LFC and stellar fibers.}
obtained during a run (Section \ref{s:obs}), since IP extraction from 
the LFC spectrum traces ``relative" variations in time.

We split the observed spectrum of each echelle order into 19 small spectral segments, each
spanning a wavelength range of $\Delta \lambda = 0.7- 1$ nm,
and the RV fitting is performed for each segment (step ``C." in Figure \ref{fig:chart}). 
Equation (\ref{eq:2}) is somewhat similar to the expression for the forward-modeling 
RV measurements in the visible using the iodine cell 
\citep[e.g., Equation (1) of ][]{2002PASJ...54..873S}. 
Also in the NIR, \citet{2010ApJ...713..410B} developed a forward-modeling technique, similar to our method, for RV measurements with the ``ammonia" cell from $K-$band spectra. 
Unlike those forward-modeling techniques, however, IPs in our RV analysis (Equation (\ref{eq:2})) are independently determined based on the reference (LFC) spectrum. 
Note that time-variable telluric lines are simultaneously modeled in our RV fit, 
which are usually ignored in the optical RV analysis. 
Below, we describe how we extract (generate) each component that appears in Equation (\ref{eq:2}), and the procedure to compute the RV value for each IRD frame.

\subsubsection{Extraction of the IP for Each Segment}

Estimation of the instantaneous IP for each frame, for each spectral segment, is the essential part of the forward-modeling technique
for precision RV measurements. In the pipelines of optical RV measurements using the iodine cell, 
instantaneous IPs are estimated from the shapes of the iodine absorption lines, which are 
blended with stellar spectrum \citep[e.g.,][]{1996PASP..108..500B, 2002PASJ...54..873S}. 
In most cases, 
the IPs are modeled by a linear combination of multiple Gaussian functions, whose heights 
are optimized simultaneously in fitting the stellar RV. 
For the case of the simultaneous reference method using LFC, the shapes of LFC emission 
lines directly reflect the instantaneous IP for each spectral segment, and we can determine the IP shape separately from the stellar RV measurement. 

Since the intrinsic width of LFC emission lines ($\sim 1\,\mathrm{MHz} = 1-2$ m s$^{-1}$
in the velocity domain) is three orders of magnitude smaller than the spectral resolution of IRD ($\approx 4$ km s$^{-1}$), 
an intrinsic LFC spectrum is well approximated by a combination of the Dirac delta functions, separated by
a fixed interval (12.5 GHz) in the frequency domain
(i.e., what we observe in the LFC spectrum is almost equivalent to the IP of the spectrograph).  
In this case, we can use the least square deconvolution (LSD) technique to extract the ``mean" line
profile for each spectral segment. In the standard LSD \citep[e.g.,][]{1997MNRAS.291..658D}, 
the observed flux vector $\mathbf{Y}$ of the LFC spectrum is expressed by a product of the 
emission-line matrix $M$ and IP vector:
\begin{eqnarray}
\label{eq:LSD}
\mathbf{Y} = M \mathrm{IP}. 
\end{eqnarray}
The matrix elements of $M$ are given by Equations (17) and (18) in \citet{1997MNRAS.291..658D}. 
For a given $M$, the least-squares solution for the IP vector is computed by 
\begin{eqnarray}
\label{eq:LSD2}
\mathrm{IP} & = &(M^\mathrm{T} S^2 M)^{-1} M^\mathrm{T} S^2 \mathbf{Y},
\end{eqnarray}
where
\begin{eqnarray}
\label{eq:LSD3}
 S^2&=& \mathrm{diag}\left( \frac{1}{\mathbf{\boldsymbol \sigma}^2}\right)
\end{eqnarray}
is the vector containing the reciprocal flux errors ($\{1/\sigma_i\}$). 
This IP extraction is easy to execute, but the IP estimation based on this standard LSD 
is known to have the ``noise amplification" problem especially when 
the input spectrum has a relatively large noise, or the pixel sampling of the input spectrum
is sparse, due to the nature of direct ``deconvolution" processes \citep{1997MNRAS.291..658D}. 
For the case of IRD's LFC, each LFC line is sampled with
only $4-5$ points (Figure \ref{fig:comb}), which are much sparser than the velocity resolution 
we want for the extracted IP ($\lesssim 0.5$ km s$^{-1}$).

Fortunately, we know that IP is generally a ``smooth" function in the velocity 
(or wavelength) domain since it originates from the spectrograph's point spread function 
on the detector. 
To take into account this {\it a priori} information, we estimate the IP by the 
Bayesian inference technique, following the formulation by \citet{2015A&A...583A..51A};
they model the pixel-to-pixel correlations in the LSD profile by a Gaussian process regression
\citep[e.g.,][]{2012MNRAS.419.2683G, 2015MNRAS.451..680E}:
\begin{eqnarray}
\label{eq:LSD3.5}
p(\mathrm{IP}|\mathbf{\boldsymbol\alpha}) = \mathcal{N}(\mathrm{IP}|\mathbf{0},~K(\mathbf{\boldsymbol\alpha})),
\end{eqnarray}
where $\mathcal{N}$ represents a multivariate Gaussian function 
and $K(\mathbf{\boldsymbol\alpha})$ is the
covariant matrix determining the pixel-to-pixel correlations in the IP, which is characterized 
by $\mathbf{\boldsymbol\alpha}$ (a vector whose dimension corresponds to the 
number of hyperparameters). 
Here, we adopt the squared exponential kernel for the covariant matrix whose components are
expressed as 
\begin{eqnarray}
\label{eq:cov}
K_{ij} = K^2\exp\left\{ -\frac{(v_i-v_j)^2}{2L^2} \right\},
\end{eqnarray}
where $v_i$ represents $i-$th velocity component of the IP function, and
$K$ and $L$ are hyperparameters that determine the amplitude and length of correlations 
in the profile. In this formulation, ${\boldsymbol\alpha}=(K,~L)$. 
Equation (\ref{eq:cov}) includes no uncorrelated (white) noise term, since we 
require a smooth functional dependence for the IP.

What we want to learn now is the posterior distribution of IP conditioning on the observed LFC spectrum $\mathbf{Y}$, 
but this probability distribution $p(\mathrm{IP}|\mathbf{Y})$ in general cannot be derived analytically. 
A solution to this is to adopt an approximation called Type-II maximum likelihood, in which
$p(\mathrm{IP}|\mathbf{Y})$ is approximated as
\begin{eqnarray}
\label{eq:LSD4}
p(\mathrm{IP}|\mathbf{Y}) &\propto& \int p(\mathbf{Y}|\mathrm{IP})p({\boldsymbol\alpha})
p(\mathrm{IP}|{\boldsymbol\alpha})d{\boldsymbol\alpha}\nonumber\\
&\approx& p(\mathbf{Y}|\mathrm{IP}) p(\mathrm{IP}|\hat{\boldsymbol\alpha}),
\end{eqnarray}
where $\hat{\boldsymbol\alpha}$ represents the set of hyperparameters that maximizes
the posterior probability of ${\boldsymbol\alpha}$ conditioning on the data $p({\boldsymbol\alpha}|\mathbf{Y})$ \citep{2015A&A...583A..51A}. 
Since $p(\mathbf{Y}|\mathrm{IP})$ is the standard likelihood
\begin{eqnarray}
p(\mathbf{Y}|\mathrm{IP}) = \mathcal{N}(\mathbf{Y}|M\mathrm{IP}, S^{-2}),
\end{eqnarray}
Equation (\ref{eq:LSD4}) reduces to 
\begin{eqnarray}
p(\mathrm{IP}|\mathbf{Y}) &\approx& 
\mathcal{N}(\mathbf{Y}|M\mathrm{IP}, S^{-2}) 
\mathcal{N}(\mathrm{IP}|\mathbf{0},~K(\hat{\boldsymbol\alpha}))\nonumber\\
&=&\mathcal{N} (\mathrm{IP}|\mathbf{\boldsymbol\mu},~\Sigma),
\end{eqnarray}
where 
\begin{eqnarray}
\Sigma &=& \displaystyle\left[ K(\hat{\boldsymbol\alpha})^{-1}+M^TS^2M\right]^{-1} \\
{\boldsymbol\mu} &=& \Sigma M^TS^2\mathbf{Y}.\label{eq:mu}
\end{eqnarray}
Equation (\ref{eq:mu}) gives the mean function of the IP based on the Bayesian LSD. 
In the absence of pixel-to-pixel correlations ($K({\boldsymbol\alpha})=0$), 
this expression is equivalent to the standard LSD solution (Equation \ref{eq:LSD2}).

The posterior probability $p({\boldsymbol\alpha}|\mathbf{Y})$ is computed as
\begin{eqnarray}
p({\boldsymbol\alpha}|\mathbf{Y})&\propto& \int p(\mathbf{Y}|\mathrm{IP})
p(\mathrm{IP}|{\boldsymbol\alpha})p({\boldsymbol\alpha})d{\mathrm{IP}}\nonumber\\
&=&p(\mathbf{\boldsymbol\alpha}) 
\mathcal{N} (\mathbf{Y}|\mathbf{0},~\Sigma_{\boldsymbol\alpha}),\label{eq:posterior}
\end{eqnarray}
where 
\begin{eqnarray}
\Sigma_{\boldsymbol\alpha} = S^2+MK({\boldsymbol\alpha})M^T,
\end{eqnarray}
and $p(\mathbf{\boldsymbol\alpha})$ is the prior distribution for ${\boldsymbol\alpha}$. 
In case that no prior is imposed on ${\boldsymbol\alpha}$, 
$\hat{\boldsymbol\alpha}$ that gives the maximum posterior probability (\ref{eq:posterior}) is 
estimated by minimizing the following $\chi^2$ statistics: 
\begin{eqnarray}
\label{eq:chisq}
\chi^2 = \mathbf{Y}^\mathrm{T} \Sigma_{\boldsymbol\alpha}^{-1}\mathbf{Y} +\log |\Sigma_{\boldsymbol\alpha}|,
\end{eqnarray}
where $|\Sigma_{\boldsymbol\alpha}|$ is the determinant of $\Sigma_{\boldsymbol\alpha}$. 


\begin{figure}
\includegraphics[width=8.5cm]{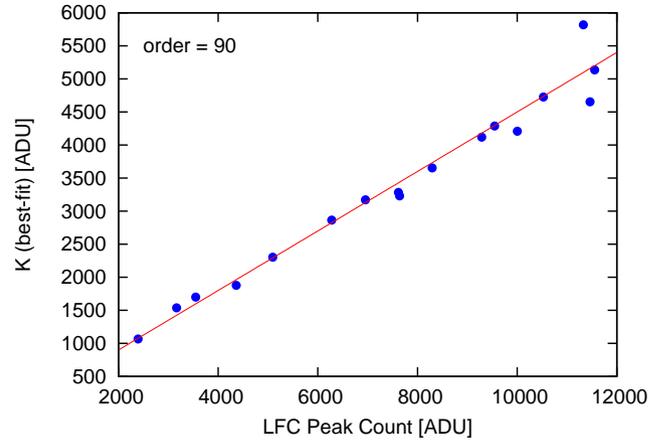}
\caption{
Optimized hyperparameter $K$ (correlation amplitude) for each spectral segment as a function of LFC's peak count within the same segment. 
}
\label{fig:GPamp}
\end{figure}
\begin{figure}
\includegraphics[width=8.5cm]{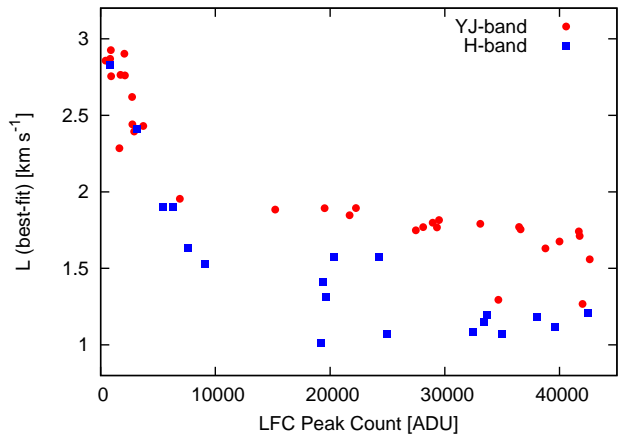}
\caption{
Optimized hyperparameter $L$ (correlation length in the velocity domain) 
as a function of LFC's meadian peak count for each order. 
}
\label{fig:GPlength}
\end{figure}
We apply this Bayesian LSD to LFC spectra and extract instantaneous IPs for individual frames. 
As we stated in Section \ref{s:IRD}, IPs of the IRD spectrograph are different
from order to order, and from segment to segment. Therefore, we split each echelle order 
of a whole LFC spectrum into 19 segments, each spanning $100-110$ pixels 
($\Delta \lambda = 0.7- 1$ nm), and computed Equation (\ref{eq:mu}) for each segment. 
This number of segments for each order was empirically determined taking into account (1) the number of available LFC emission lines, 
(2) similarity/difference in the IPs of adjacent segments, and (3) CPU time to compute the inverse matrix in the LSD. 
Spectral segments corresponding to both edges of each order are excluded from the analysis, since 
significant fractions of pixels for those segments have near-zero flux counts due to vignetting on the detector.

Ideally, one should optimize the hyperparameter ${\boldsymbol\alpha}$ for each frame, 
for each spectral segment, by minimizing Equation (\ref{eq:chisq}). However, the equation 
involves an inverse matrix calculation, which is computationally expensive. In the real 
data analysis, therefore, we decided to take an approach to fix the hyperparameters in computing 
Equation (\ref{eq:mu}) to empirical values pre-determined by our analysis of typical LFC spectra;
the implicit assumption in this empirical approach is that the hyperparameters that give
the best description of observed LFC spectra do not significantly vary for each frame, 
but depend only on the peak counts and wavelengths of LFC emission lines.

Allowing the two hyperparameters in Equation (\ref{eq:cov}) 
to float freely, we performed the minimization of Equation (\ref{eq:chisq}) 
for a set of typical LFC spectra using the Nelder-Mead simplex method \citep[e.g.,][]{2002nrc..book.....P}. 
Figure \ref{fig:GPamp} plots an example of the optimized hyperparameter $K$ for 17 
segments of one specific echelle order. The horizontal axis of the figure is the mean peak 
count of LFC 
emission lines within the same segment. As is evident from the figure, $K$ giving the 
minimum $\chi^2$ is almost proportional to the mean peak of the LFC lines. The red solid 
line in the same figure indicates the result of a linear regression to the 17 points 
(slope $\approx 0.45$). We also checked for the variation of the optimal $L$ (the correlation scale in the velocity domain) for many different segments, 
finding that the optimal $L$ is also dependent on the typical LFC intensity 
(i.e., S/N ratio) for individual orders. 
The optimal $L$ averaged within each order is plotted in Figure \ref{fig:GPlength} 
as a function of the median LFC peak count of that order; 
A higher value of $L$ is preferred for orders with lower LFC intensities, 
while the quantitative behaviors are slightly different for $YJ-$band and $H-$band
detectors. The velocity scale covered by each pixel of the detector differs
significantly from order to order (and from segment to segment), but this general trend 
implies that the optimal correlation length is grossly affected by the S/N ratio of
the LFC lines, and a longer correlation length is generally required to smooth out 
noisier spectra. 
From those analyses, we derived the empirical values for the two hyperparameters for 
each order, and those parameters are held fixed at those values in the subsequent analyses.

\begin{figure}
\includegraphics[width=8.5cm]{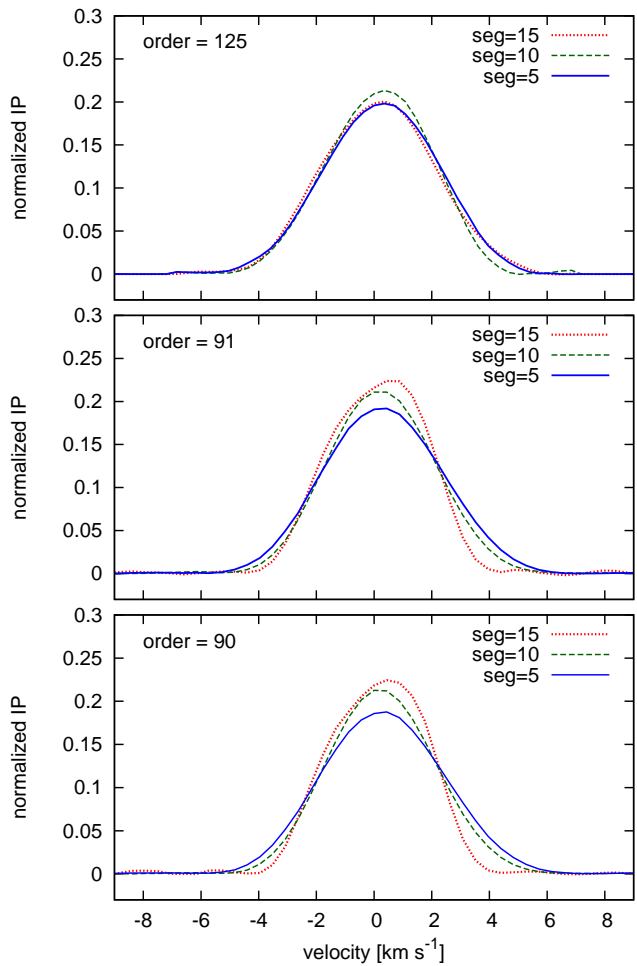}
\caption{Extracted IPs for three different spectral segments of echelle order 125 (top), 91 (middle), and 90 (bottom). 
}
\label{fig:IP}
\end{figure}

Figure \ref{fig:IP} plots instances of the IPs estimated using Equation (\ref{eq:mu})
for an observed LFC spectrum, taken on UT 2018 August 6. 
The three panels show the three different segments' IPs for echelle $\mathrm{order}=125$ 
($1171\,\mathrm{nm}<\lambda<1185\,\mathrm{nm}$), $\mathrm{order}=91$ 
($1608\,\mathrm{nm}<\lambda<1629\,\mathrm{nm}$),
and $\mathrm{order}=90$ ($1626\,\mathrm{nm}<\lambda<1647\,\mathrm{nm}$), respectively. 
As expected, the IP shape in the velocity domain is heavily dependent on the position of the detector 
and each IP is not symmetric with respect to the line center. 
Moreover, the IP variation within one order is not always monotonic against wavelength;
the IP is the sharpest around the central part (segment = 10) of the 125th spectral order, while
they are similar in shape in both edges (segment = 5 and 15) of the spectrum for that order. 
On the other hand, IP has the highest peak at longer wavelengths for $\mathrm{order}=90$
and $\mathrm{order}=91$. 
IPs of the same segment number in the neighboring orders
are similar in shape, as shown in Figure \ref{fig:IP}.

IPs are well extracted only when the S/N ratio of the spectral segment is sufficiently 
high enough, but it is not straightforward to estimate IPs for segments with low S/N ratios. 
In particular, when a LFC spectrum is taken with relatively short integration time ($<1$ minute), 
the detector's readout noise is more significant in comparison with LFC emissions in 
some orders \citep[cf.][]{2018SPIE10702E..60K}. 
In addition, as a characteristic of IRD's LFC, emission lines are not generated with good 
S/N ratios at certain wavelengths 
(e.g., around 1100 nm and 1550 nm; see Figure \ref{fig:comb}).  
For those segments, we estimate the IPs by interpolating the IPs of segments that are 
neighboring, for which IPs are properly extracted from the LFC emission lines. 
The interpolation does not only use the adjacent (or near) segments
within the same order, but also refers the IPs of similar segment numbers in the 
neighboring orders, on the assumption that the point spread function 
of the spectrograph ``gradually" varies in both spatial and wavelength directions on 
the detector. 
The similarity of IPs in the neighboring orders shown in Figure \ref{fig:IP} reinforces this statement. 
Since the current LFC covers wavelengths between 1050 nm and 1730 nm, 
we can only extract IPs for the segments beyond 1050 nm. Below 1050 nm, we adopt the mean 
IPs over the $Y$-band spectrum, but those segments below
1050 nm are not used in the RV analysis for observed spectra (Section \ref{s:onsky}).

\subsubsection{Telluric Absorption Spectrum $T(\mathbf{A};\lambda)$}\label{s:lblrtm}
For the telluric transmittance $T(\mathbf{A};\lambda)$, we use the theoretical 
transmission spectra generated by Line By Line Radiative Transfer Model 
\citep[LBLRTM:][]{2005JQSRT..91..233C}. 
To save the computation time in fitting the spectrum, we synthesized the telluric transmittance 
at $4\times 4$ grid points for the precipitable water vapor content ($W$) and target's 
airmass ($A$), covering realistic ranges of those parameters 
($1.0\,\mathrm{mm}<W<5.0\,\mathrm{mm}$ and $1.0<A<2.9$). 
For each grid point in the $\mathbf{A}=(W, A)$ plane, we generated 
a telluric spectrum for the atmosphere above the summit of Maunakea. 
In doing so, the $T-p$ (and height $H$) profile and volume mixing ratio 
of each atmospheric molecule are required. 
Following \citet{2016A&A...585A.113R}, we employed 
the averaged profiles based on the 
Global Data Assimilation System (GDAS)\footnote{https://ready.arl.noaa.gov/READYcmet.php} 
sounding files at the location of Maunakea for $H\leq 26$~km, and also downloaded 
the MIPAS\footnote{The Michelson Interferometer for Passive Atmospheric Sounding (MIPAS) model 
atmosphere: 
http://www-atm.physics.ox.ac.uk/
}
mid-latitude night-time profiles for $H>26$ km. We then input
these profiles into the LBLRTM code, which generates the 
telluric transmission spectrum in the NIR with the input variables, $(W, A)$. 
In the RV fit (Equation \ref{eq:2}), the telluric spectrum $T(\mathbf{A};\lambda)$
is generated by interpolation of those template telluric spectra on the $(W, A)$ grid, 
and $\mathbf{A}$ is optimized simultaneously. 

In the NIR, some spectral segments are also contaminated by night-glow emission lines, which are particularly prominent in the $H$-band data. In the RV analysis, 
we simply mask all those emission lines based on the theoretical radiance model generated by \texttt{SkyCalc} \citep{2012A&A...543A..92N, 2013A&A...560A..91J}, by which the number of usable pixels for spectrum fitting is reduced by only $1-2\%$ for most echelle orders.

\subsubsection{Estimation of the Intrinsic Stellar Spectrum $S(\lambda)$ (Template for RV Fits)}

One tricky part of the forward-modeling technique for RV measurements in the NIR is 
the estimation of the intrinsic stellar spectrum $S(\lambda)$. 
In the NIR, stellar spectra are heavily contaminated by the telluric absorptions 
and nightglow emissions, which complicates the extraction of a telluric-free
stellar template. Moreover, the molecular line lists are often incomplete or inaccurate
\citep[e.g.,][]{2018Atoms...6...26T}, and thus theoretically synthesized
spectra disagree with observed ones, meaning that we cannot use those model spectra
as templates for RV measurements by the forward-modeling technique.

Fortunately though, stellar line positions are not constant in time due to the barycentric
motion of Earth, while telluric line positions are almost unchanged against wavelength, 
which helps us disentangle the stellar lines from telluric ones. 
To extract the stellar template, the first step is deconvolution of the instantaneous 
IP for each spectral segment (step ``B1." in Figure \ref{fig:chart}). This is carried out by using the IP 
estimated from the LFC spectrum for the same spectral segment.  
For IP deconvolutions, we use the ``iterative/recursive" deconvolution described in \citet{1994rhis.conf...24C}
and \citet{2002PASJ...54..873S}, in which, unlike LSD, one does not need to assume the intrinsic profile 
is expressed by the delta function.

The second step is to remove the telluric absorption lines 
from the IP-deconvolved spectrum (step ``B2." in Figure \ref{fig:chart}) using 
theoretically synthesized telluric spectra as in Section \ref{s:lblrtm}. 
We apply the least-squares technique to each echelle order of an observed spectrum 
in order to model the telluric transmittance and estimate the best-fit telluric parameters ($\mathbf{A}$ and $v_\mathrm{tel}$), together with the continuum polynomial.
Since intrinsic stellar lines are not known at this point and they are blended with 
telluric features in each observed spectrum, fitting the telluric lines by theoretical 
models is affected by the contaminating stellar lines. To mitigate this impact, 
we empirically 
imposed more ``weights" on pixels corresponding to deeper telluric lines so that spectral parts having no (or very shallow) telluric features have minimal contributions to the spectrum fitting.

We developed another option to remove telluric absorptions to create an intrinsic stellar 
spectrum S($\lambda$). This is based on the telluric standard star (rapid rotator having a
featureless spectrum) immediately observed before or after an RV target. 
By dividing the target spectrum by a normalized spectrum of the telluric standard, 
we can remove telluric absorption lines, unless they are saturated (i.e., near zero flux counts). 
Note that these observed telluric spectra are used only for the construction of the 
stellar template, and not each time for an RV measurement.
For this option, we deconvolve the IP from the telluric-removed stellar spectrum 
to obtain the intrinsic stellar template. 
This operation is not mathematically equivalent to the procedure for real data acquisition; 
multiplication of the telluric transmittance and convolution of IP are not mathematically
commutative. However, our experience with the observed data suggested the deconvolution 
``after" the telluric removal yields a good approximation to the intrinsic stellar template. 
This procedure also does not remove the nightglow emission lines.

\begin{figure}
\includegraphics[width=8.5cm]{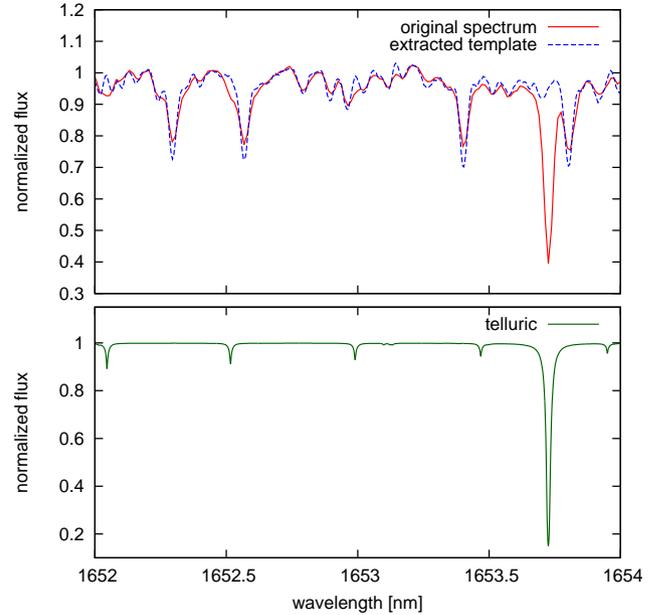}
\caption{
{\it (Upper)} Observed spectrum (before any processing; red solid line) v.s. IP-deconvolved, telluric-removed template (blue dashed line)
for GJ 699. 
{\it (Lower)} Theoretical telluric transmittance in the same region, before the convolution of the IP.  
}
\label{fig:extract_temp}
\end{figure}
Each IP-deconvolved, telluric-removed stellar spectrum derived by the above steps is cross-correlated against a theoretical 
stellar template \citep[PHOENIX BT-SETTL:][]{2013MSAIS..24..128A} to roughly estimate the 
stellar RV for that frame, and the spectrum is Doppler-shifted by the RV such that the 
resulting spectrum is in the stellar rest frame. 
The theoretical template for cross-correlations should ideally be generated based on 
the accurate stellar parameters of the target star (e.g., $T_\mathrm{eff}, \log g, [\mathrm{Fe/H}]$), 
but our experience has shown that a small difference in those parameters has a 
negligible impact on the rough RV estimation. We thus prepared only two theoretical 
templates for RV measurements of our target stars presented in this paper
as well as other M-dwarf targets during the engineering observations
($T_\mathrm{eff}=2700, 3100$ K, $\log g=5.0$, $[\mathrm{Fe/H}]=0.0$). 
Since deconvolution is known to increase 
the flux noise and telluric removal by the above steps cannot completely 
clean off telluric lines as well as nightglow emissions for individual frames, 
we median-combine multiple frames ($>10$ preferred) to gain a high S/N stellar template, 
free of telluric lines. In doing so, it is important to obtain spectra of the same target 
star on well separated nights; the stellar lines can be Doppler-shifted by Earth's 
barycentric motions by up to $\approx \pm 30$ km s$^{-1}$ (except targets at 
high ecliptic latitudes). 
Thus, acquisitions of multiple-epoch spectra whose barycentric RV corrections are separated by $\gtrsim 4$ km s$^{-1}$ 
(= resolution of IRD) enable us to well distinguish stellar line positions from the telluric ones.

Figure \ref{fig:extract_temp} depicts a comparison between an observed IRD spectrum of 
GJ 699 and the extracted template for the same target based on the above procedure. 
Both spectra are normalized and Doppler-shifted to the same reference. 
More than 20 frames are combined to obtain this template. 
As is evident in the figure, stellar lines become sharper by deconvolution of IP. 
The strong line at $\approx 1653.7$ nm is telluric absorption, which was completely 
removed in the output stellar template.

\begin{figure*}
\begin{center}
\includegraphics[width=15cm]{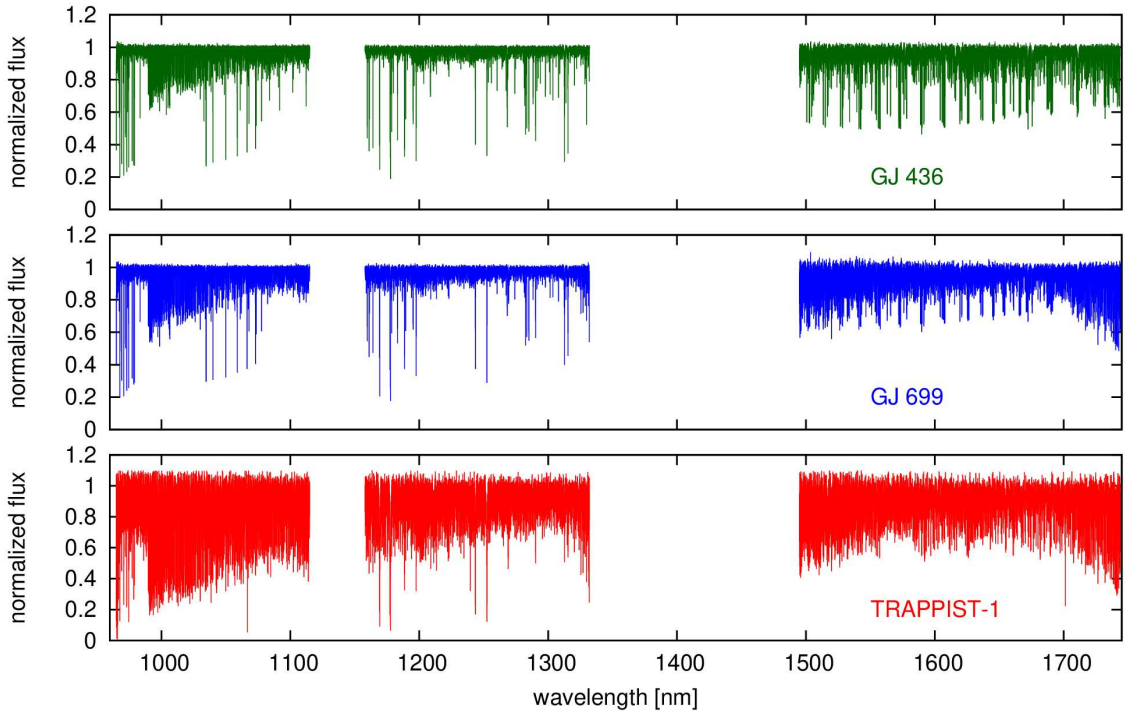}
\end{center}
\caption{Comparison of the extracted template spectra for three different M dwarfs (GJ 436, GJ 699, and TRAPPIST-1 from top
to bottom). The absorption features, between 990 nm and 1050 nm in particular, become deeper 
for later spectral types. 
}
\label{fig:templates}
\end{figure*}
We applied our pipeline to extract intrinsic stellar templates for various stars observed 
by IRD. Figure \ref{fig:templates} illustrates three examples of extracted stellar templates 
(deconvolved and telluric-removed): an early-M dwarf (GJ 436: top), a mid-M dwarf 
(GJ 699: middle), and a late-M dwarf (TRAPPIST-1: bottom). 
The two gaps at $1110-1150$ nm and $1330-1495$ nm correspond to the strong telluric regions and the gap between 2 detectors, respectively, 
for which we are unable to extract clean templates.  
Figure \ref{fig:templates} indicates that the later-type stars have deeper and denser features of molecular lines, 
which leads to higher RV information content for those targets.

\subsubsection{RV Fitting for Individual Segments}

\begin{figure}
\includegraphics[width=8.5cm]{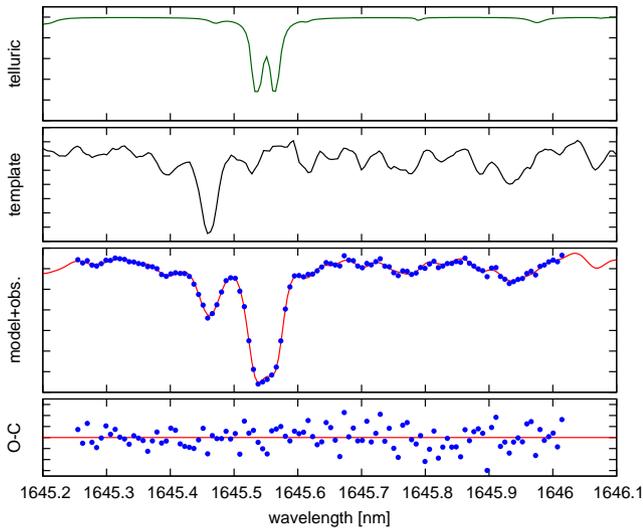}
\caption{
Spectrum fitting by our pipeline for a segment in the $H-$band (GJ 699). From top to bottom, the best-fit theoretical telluric 
transmittance $T(\mathbf{A};\lambda)$, stellar template $S(\lambda)$, 
observed spectrum (plus the best-fit model in red), $O-C$ residuals are plotted, respectively. 
}
\label{fig:rvfit}
\end{figure}
Using all the components generated by the above steps, 
we fit the observed spectrum $f_\mathrm{obs}(\lambda)$ by Equation (\ref{eq:2}) for each 
spectral segment spanning $\approx 1$ nm. 
The basic fitting parameters here are the three coefficients of the polynomial $k(\lambda)$,
$\mathbf{A}$, $v_\star$, and $v_{\rm tel}$ (seven parameters in total). 
Note that telluric transmittance is simultaneously modeled again when RVs are computed by 
fitting the template spectrum to individual observed spectra.
In fitting each spectral segment, IP is fixed to the one estimated from the corresponding 
segment of the simultaneously taken LFC spectrum. 
The optimization of the fitting parameters is performed with the Levenberg–Marquardt (LM) $\chi^2$ minimization technique 
\citep[e.g.,][]{2002nrc..book.....P}, or our customized code of the Markov Chain Monte Carlo (MCMC) samplings
\citep[e.g.,][]{2015ApJ...799....9H}; 
the LM optimization is about twice as fast as the MCMC analysis. 
For the latter method, we can optionally impose Gaussian priors on some fitting parameters,
(e.g., telluric parameters $\mathbf{A}$ and $v_\mathrm{tel}$). 
Through the analyses of observed spectra, we confirmed that these two optimizations give almost equivalent 
RV results for nominal targets (slowly rotating M dwarfs), but for stars with moderate rotations ($v\sin i>5$ km s$^{-1}$)
RV results behaved slightly better in the case of the MCMC fitting by imposing telluric priors. 
This investigation is under progress, and we hope to present the result in future works. 
For the analyses of observed spectra (Section \ref{s:onsky}), we present the RV results based on the MCMC analyses. 
Figure~\ref{fig:rvfit} depicts an example of our fitting procedure for a segment
($\mathrm{order}=90$, $\mathrm{part}=17$) of an observed IRD spectrum of GJ 699 (blue points).

A total of $\approx 1000$ segments, from $970$ nm to $1700$ nm, are analyzed for RV measurements, 
though a significant fraction ($>40\%$) of those segments are not usable for RV measurements 
due to low blaze efficiency around edges of each echelle order and/or very strong 
telluric lines. Currently LFC spectra only cover between $1050-1720$ nm, thus we are not 
capable of extracting accurate IPs for spectral segments below $\approx 1050$ nm. 
The RV error for each segment is estimated based on the covariant matrix for the LM $\chi^2$
minimization, or the marginalized posterior distribution of $v_\star$ for the MCMC analysis. 
The final RV and its uncertainty are determined from the weighted mean of the RV values 
for all the available (converged) segments after clipping out the segments showing 
unfavorable behaviors; 
Segments with an imperfect removal of telluric lines in $S(\lambda)$ and/or bad estimations of 
IPs produce anomalistic RV behaviors in comparison with those of the neighboring segments. 
Those segments are therefore removed in computing the weighted mean for $v_\star$.

It should be emphasized that our methodology described above relies on 
the critical assumption that the IP extracted from the LFC spectrum (from the 
reference fiber) is identical to that of the stellar spectrum (from the 
stellar fiber); albeit the two fibers (both multi-mode fibers) are identical in shape (circular) and diameter, there is no guarantee that the two fibers produce
exactly the same point-spread function on different positions of the detector. 
One test to verify this assumption is to implement a laboratory
experiment with the LFC being injected into both fibers, and check for the magnitude
of the ``relative" temporal drift between the two fibers. The result of this experiment is
presented in detail in \citet{2018SPIE10702E..60K}, in which we demonstrated 
that the relative
RV variation between the two LFC spectra is $1.3-1.9$ m s$^{-1}$ including the 
random Poisson plus readout noise over a time scale of $\approx 2$ weeks
(see Figure 3 of \citet{2018SPIE10702E..60K}). 
In addition to the fiber-induced difference in IPs, one also needs to account for
the different paths which the stellar light and LFC light pass through in the case
of on-sky observations; time-variable AO corrections coupled with limited fiber
scrambling can lead to a variation in IPs only for stellar spectra, potentially 
resulting in an apparent shift in stellar RVs. 
All these concerns motivated us to conduct on-sky observations of several RV
standard stars with IRD to quantify the impact of differing IPs for the two
IRD fibers. The on-sky stability of stellar RVs is presented in Section \ref{s:onsky}.


\section{Validation of the RV Pipeline Based on Mock Spectra}\label{s:mock}

\subsection{Setups}\label{s:setups}

As the first test of our RV analysis pipeline, we estimated ``theoretical" RV precisions 
that IRD can achieve. To do so, we performed a series of Monte Carlo simulations using 
theoretical model spectra in the NIR. 
Although this sort of numerical simulations have been carried out in 
the past \citep[e.g.,][]{2010ApJ...710..432R, 2011A&A...532A..31R}, we repeat similar simulations 
to take into account IRD's specifications such as the wavelength coverage, 
spectral resolution, pixel sampling, wavelength-dependent instrumental efficiency, etc. 
Below, we briefly describe our Monte Carlo
simulations to estimate IRD's RV precisions for each spectral type. 
The precisions derived here are the uncertainties in $v_\star$ of Equation (\ref{eq:2}), 
originating from the Poisson noise (and readout noise) in the stellar spectra as well as 
the pipeline capability to fit the NIR spectra in the presence of blending telluric lines. 
In this section, we do not take into account the imperfect removal of telluric lines in the 
stellar template $S(\lambda)$ and imperfect estimation of IPs for individual segments. 
The RV analysis for the actual observed spectra will be presented in Section \ref{s:onsky}.

We began with a theoretical NIR spectrum generated 
by the BT-SETTL model \citep{2013MSAIS..24..128A}. 
Here, we adopt three types of M dwarfs and one solar analog: a late M 
($T_\mathrm{eff}=2500$ K), a mid-M ($T_\mathrm{eff}=3000$ K), an
early M ($T_\mathrm{eff}=3500$ K) dwarfs, and a G dwarf ($T_\mathrm{eff}=5800$ K). 
We set the metallicity to $[\mathrm{Fe/H}]=0.0$ for the fiducial case, but we will later use 
templates with non-solar metallicities to investigate the metallicity dependence of 
RV precision (Section \ref{s:metal}).
Since RV precisions are known to be highly dependent on stellar rotation ($v\sin i$), 
we convolved each PHOENIX spectrum with the rotation plus macroturbulence
broadening kernel, following \citet{2011ApJ...742...69H}. 
For the macroturbulent velocity $\zeta$ in the radial-tangential model \citep{2005oasp.book.....G}, 
we adopted $\zeta=1$ km s$^{-1}$ for M dwarfs following \citet{1998ApJ...498..851V} and \citet{2006ApJ...652.1604B},
and $\zeta=3.98$ km s$^{-1}$ for the solar analog \citep{2005ApJS..159..141V}. 
We then multiplied the broadened spectrum by the telluric transmission 
spectrum synthesized by LBLRTM for an arbitrary observing condition on Maunakea
(we will also use the observed telluric transmittance in Section \ref{s:telluric}). 
To simulate IRD observations, this theoretical stellar plus telluric spectrum was multiplied 
by IRD’s wavelength-dependent efficiency, and convolved with the corresponding IP for 
each segment. The efficiency is based on our observations of a rapid 
rotator\footnote{http://ird.mtk.nao.ac.jp/IRDpub/index\_tmp.html}, reflecting the 
blaze function for each echelle order, fiber transmittance, AO efficiency, detector 
quantum efficiency, etc. For simplicity, we fixed the IPs for individual segments to 
the observed ones (extracted by LSD) on an arbitrary night. 
Finally, the resulting spectral fluxes were converted into the photon counts
and stored with exactly the same pixel sampling as IRD's detector. The Poisson noise was 
estimated for each pixel by scaling the S/N ratio to the value at a reference wavelength, for which we adopted $1000$ nm. 
We also took into account the readout noise by adding white noise of 45 e$^-$ per pixel. 
This noise level approximately corresponds to the readout noise for a 1-minute integration 
\citep[i.e., the nominal integration time for GJ 699;][]{2018SPIE10702E..60K}.

The rest of this section presents the simulated results of mock RV analyses, in which we 
put the mock IRD spectra generated by the above steps into our RV pipeline.  
Here, the ideal RV precisions of the IRD spectrograph are estimated by focusing on the 
spectrum fitting procedure in step ``C." of Figure \ref{fig:chart}. 
We set the IPs of individual segments to the ones used in creating mock IRD spectra. 
The adopted stellar template in the RV fit is identical to the input one under the assumption 
that telluric lines are perfectly removed via the process of extracting the stellar template 
(step ``B." in Figure \ref{fig:chart}) using a large number of spectra.
We allowed all the seven fitting parameters (including telluric parameters) to vary in the fit.

\begin{figure}
\includegraphics[width=8.5cm]{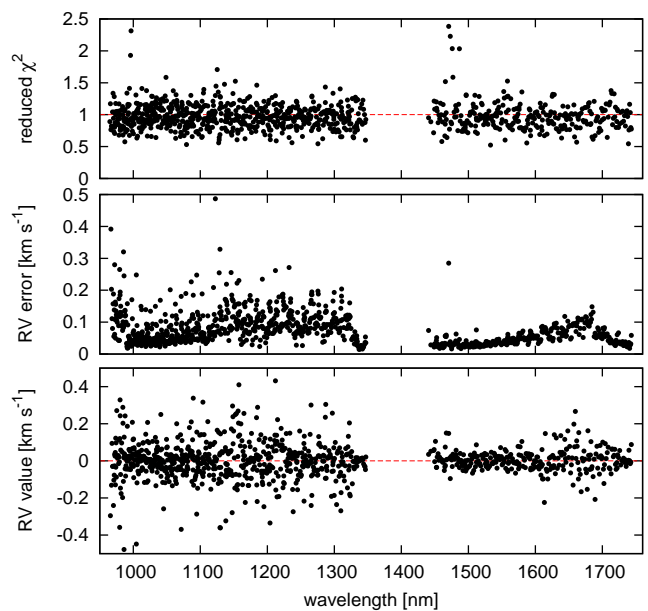}
\caption{
Sample result of the RV analysis for a mock IRD spectrum 
(fiducial case: $T_\mathrm{eff}=3000$ K, $v\sin i=1$ km s$^{-1}$, $\mathrm{[Fe/H]}=0.0$, $\mathrm{S/N}\approx 100$ at 1000 nm). 
RV values (relative to the template), their errors, and reduced $\chi^2$ are plotted for individual segments, from bottom to top, respectively. 
}
\label{fig:simu1}
\end{figure}
Figure \ref{fig:simu1} plots an example of the fitting result for a mock IRD spectrum 
with $T_\mathrm{eff}=3000$ K, $v\sin i=1$ km s$^{-1}$, $[\mathrm{Fe/H}]=0.0$, 
and $\mathrm{S/N}\approx 100$ per pixel at 1000 nm, 
which we call the ``fiducial case" in the subsequent analyses. 
The best-fit RV value, its statistical error, and the reduced $\chi^2$ value for the fit of each spectral segment
are plotted as a function of the central wavelength of the segment, from bottom to top,
respectively. The absence of data points between 1350 nm and 1450 nm corresponds to the 
gap between the two IRD detectors. 
As shown in the figure, the RV precision for each segment is generally better in the $H$-band, 
partly due to higher S/N ratios at longer wavelengths, but the number of segments for the RV fit
is larger in the $Y+J$ bands. 
Figure \ref{fig:simu1} includes the fitting results for spectral segments between $Y$ and 
$J$ bands ($1113-1160$ nm), but since this spectral region is heavily contaminated by 
strong telluric absorptions, those segments are not used to compute the overall RV precision 
for each frame. 
The reduced $\chi^2$ between the mock spectrum and best-fit model is distributed around 1.0, 
suggesting a good fit to each segment. 
A small number of segments ($\approx 10$) show relatively large $\chi^2$ values ($\gtrsim 1.5$). 
Those segments are found to have especially strong telluric absorptions, which most likely 
led to the parameters (and $\chi^2$) captured at a local minimum. 
Such segments showing anomalies are clipped and ignored in deriving the final RV value
from all the segments.

\subsection{Doppler-shifting the Spectra}\label{s:Doppler}

\begin{figure}
\includegraphics[width=8.5cm]{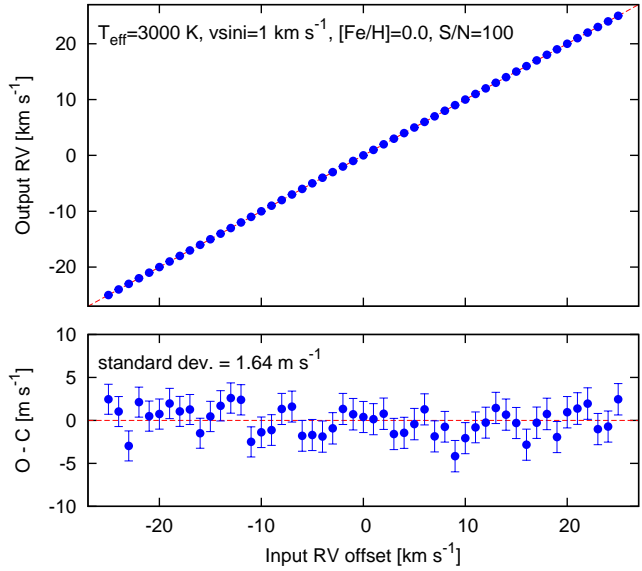}
\caption{
Input RV offset (given to the template) v.s. output RV returned by our pipeline. 
The bottom panel plots the residual between the input and output RVs in m s$^{-1}$. 
}
\label{fig:simu2}
\end{figure}
Due to the motion of the observer with respect to the barycenter of the solar system, the 
positions of stellar lines relative to the telluric ones are always shifting in time, 
implying that the magnitude of blending between stellar and telluric lines is not the same 
for each spectrum. 
This time-variable blending of stellar and telluric lines leads to a different level of degeneracy 
between the stellar RV ($v_\star$) and telluric parameters in fitting each spectrum, which may cause 
systematically large scatters in the time sequence of final RVs. 
Here, in order to check if 
\begin{enumerate}
\item the RV error returned by our analysis pipeline (i.e., the ``internal" error) is consistent with the scatter of RV points
calculated through many trials of the mock data analysis, and
\item variations of stellar line positions relative to the telluric line positions have a minor effect on the 
output RVs,
\end{enumerate}
we generated 51 mock IRD spectra for the fiducial case; We repeatedly Doppler-shifted the input stellar template
with the velocity step of $1$ km s$^{-1}$, from $-25$ km s$^{-1}$ to $+25$ km s$^{-1}$, 
simulating a situation that the same target is observed for a whole year around. 
Note that for simplicity we do not take into account the fact that the stars themselves 
can have a systemic velocity of up to about $\pm 50$ km s$^{-1}$ with respect to the Sun.

The result of analyzing the 51 spectra is shown in Figure \ref{fig:simu2}. 
In the top panel, the output RV values are plotted against the input RV offsets given to the stellar template. 
The bottom panel of Figure \ref{fig:simu2} plots the residual between the input and output RVs. 
The standard deviation of the 51 RV residuals was 1.64 m s$^{-1}$, which is almost consistent
with the mean internal error for those points (1.80 m s$^{-1}$). 
In order to investigate the impact of blending between stellar and telluric lines and resultant possible 
correlation between the input RV and the $O-C$ residual, 
we fitted the RV residual by a linear function of the input RV offset, and derived the slope $\kappa$. 
The resulting slope was found to be $\kappa=-0.019\pm 0.016$ m s$^{-1}$ (km s$^{-1}$)$^{-1}$, consistent with 
zero within about $1\,\sigma$. 
All these results imply that the spectral fitting module in our RV-analysis pipeline works well overall, 
returning RV values and their errors that are almost self-consistent.

\subsection{S/N Dependence}\label{s:snr}

\begin{figure}
\includegraphics[width=8.5cm]{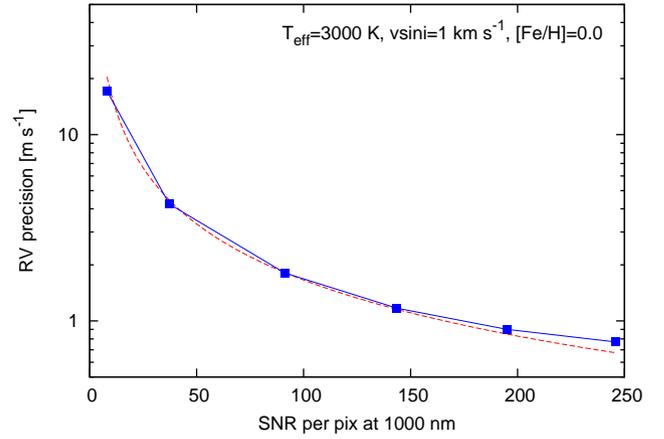}
\caption{
Simulated RV precision as a function of the nominal S/N ratio (Equation \ref{eq:SN}). 
The red dashed curve represents the best-fit regression assuming the precision is 
inversely dependent on the S/N ratio.
}
\label{fig:simu3}
\end{figure}
Next, we checked the dependence of RV precision on S/N ratio of the spectrum. 
We simulated RV measurements for the fiducial case, but varied the S/N at 1000 nm. 
We set the counts in the pixel of 1000 nm to $20^2$, $50^2$, $100^2$, $150^2$, 
$200^2$, and $250^2$ e$^-$ and scaled the flux counts and their Poisson noise for the 
other pixels. A fixed Gaussian readout noise ($\mathrm{RN}=45$ e$^-$) was added 
to each pixel, as in Section \ref{s:Doppler}. 

Figure \ref{fig:simu3} plots the internal errors for the RV measurements with different 
S/N ratios. 
The horizontal axis in Figure \ref{fig:simu3} is the ``nominal" S/N ratio computed by 
\begin{eqnarray}
\label{eq:SN}
\mathrm{S/N}_\mathrm{nominal} = \frac{F(1000~\mathrm{nm})}{\sqrt{F(1000~\mathrm{nm})+\mathrm{RN}^2}},
\end{eqnarray}
where $F(1000~\mathrm{nm})$ is the flux count in e$^-$ at 1000 nm. 
As expected, the RV precision is almost inversely proportional to the S/N;
a regression with $y=ax^{-1}$ to the result is drawn by the red dashed line in 
Figure \ref{fig:simu3}. 
At very high S/N ($\gtrsim 200$), the RV precision is slightly higher than the $y=ax^{-1}$ curve,
suggesting that we have some room for improvement in the RV analysis pipeline;
in particular, a higher-order continuum polynomial ($k(\lambda)$) and higher resolution 
in the convolution (numerical integrations) may be required in Equation (\ref{eq:2}) for 
the case of high S/N spectra.

\subsection{Dependence on Spectral Type and Rotation Velocity}\label{s:teff}

\begin{figure}
\includegraphics[width=8.5cm]{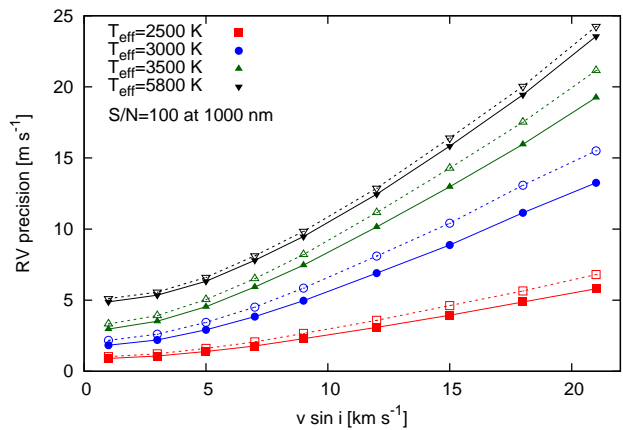}
\caption{
Simulated RV precision as a function of stellar $v\sin i$ for four different types of stars
($T_\mathrm{eff} = 2500, 3000, 3500, 5800$ K). The solid lines (filled symbols) present the 
RV precisions when IRD's whole spectrum is used, while the dashed lines (open symbols) 
correspond to those if only the wavelength range above $1050$ nm (which is covered by 
the current LFC) is used. 
}
\label{fig:simu4}
\end{figure}
Spectral type and rotation velocity of the star are the two major components that 
determine the RV precision. Setting the flux counts to $100^2$ e$^-$ at 1000 nm, we created 
a number of mock IRD spectra for differing spectral types and rotation velocity. 
We employed four different $T_\mathrm{eff}$ from the BT-SETTL model \citep{2013MSAIS..24..128A} 
as stated above, and adopted $v\sin i=1, 3, 5, 7, 9, 12, 15, 18$, and 21 km s$^{-1}$ for the rotation velocity. 
In Figure~\ref{fig:simu4}, we plot the simulated
RV precisions as a function of $v\sin i$. 
Solid lines (filled symbols) show the RV precisions when IRD's whole spectral range 
($970\,\mathrm{nm}<\lambda<1744\,\mathrm{nm}$)
is used for the RV analysis, while dashed lines (open symbols) indicate those for the
wavelengths that are currently covered by
the LFC spectrum ($1050\,\mathrm{nm}\lesssim \lambda\lesssim 1730\,\mathrm{nm}$). 
More rapidly rotating stars exhibit worse RV precision due to line-broadening, 
resulting in less Doppler information, while later-M dwarfs show better RV precisions even
for relatively large $v\sin i$. 
This is mainly because late M dwarfs generally have a larger number of molecular lines and 
deeper features in their spectra, increasing intrinsic Doppler information in the spectra.
See Figure \ref{fig:templates} regarding 
how later-type stars exhibit richer features of molecular lines.

For early M dwarfs, we need $\mathrm{S/N}\gtrsim150$ to achieve a precision of $< 2$ m s$^{-1}$, 
but similar RV precisions are achievable with only $\mathrm{S/N}=100$ for late-M dwarfs
with moderate rotations ($v\sin i<8$ km s$^{-1}$). 
The number of absorption lines is much smaller for solar-type stars, which results in
limited RV precisions in the NIR (black line in Figure \ref{fig:simu4});
as stated in the literature \citep[e.g.,][]{2010ApJ...710..432R, 2011A&A...532A..31R}, 
RV precisions for solar-type stars are much better at optical wavelengths. 
Meanwhile, RV measurements in the NIR have an advantage over optical measurements
in terms of reduced stellar jitter, especially for young stars, due to mitigated contrasts 
of active regions on the stellar surface \citep[e.g.,][]{2019RNAAS...3...89B}.

\subsection{Metallicity Dependence}\label{s:metal}

\begin{figure}
\includegraphics[width=8.5cm]{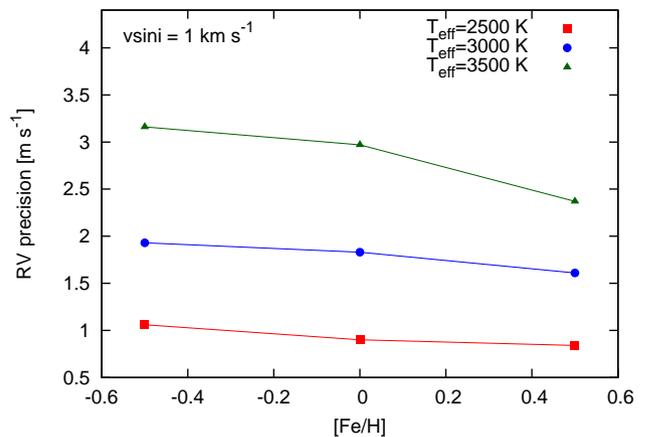}
\caption{
Simulated RV precisions as a function of [Fe/H] for the input stellar templates 
($T_\mathrm{eff} = 2500, 3000, 3500$ K, $v\sin i=1$ km s$^{-1}$, $\mathrm{S/N}\approx 100$ at 1000 nm). 
}
\label{fig:simu5}
\end{figure}
In discussing the occurrence rate of planets revealed by blind RV surveys, one should keep 
in mind that metal-rich stars have deeper absorption features at all wavelengths and the
RV precision tends to be better, facilitating the detection of planets. In order to learn 
to what extent RV precisions are improved or degraded by stellar metallicity, we repeated 
the numerical experiment for M dwarfs above with three different metallicities:
$\mathrm{[Fe/H]}=-0.5,\,0.0, \,+0.5$\footnote{Metal-rich M dwarfs with $\mathrm{[Fe/H]}>0.4$
are very rare, but we test extreme cases here.}. 
In the mock analyses, we adopted $v\sin i=1$ km s$^{-1}$
and set the flux count to $100^2$ e$^-$ at 1000 nm as in the fiducial case. 
The result ($[\mathrm{Fe/H}]$ v.s. RV precision) is shown in Figure \ref{fig:simu5}. 
RV precisions were found to be better (worse) for metal-rich (metal-poor) M dwarfs than
solar-metallicity stars by $5-20\,\%$. 
One needs to account for the metallicity dependence of RV precision in implementing
planet yield simulations for blind Doppler surveys, but our simulated result 
suggests that the RV precision depends more on the spectral type of the target 
(i.e., the number of lines) than the metallicity.

\subsection{Impact of Telluric Lines}\label{s:telluric}

In the mock RV analyses above, we used the theoretical telluric transmittance both in 
creating mock spectra and fitting the mock data, which corresponds to the ideal cases of 
our RV measurements with the best achievable precisions.
In reality, however, theoretical telluric spectra can disagree with the actual (observed) 
telluric ones due to incomplete molecular line lists, imperfect input atmospheric profiles, and 
breaking of approximations used in the theoretical calculations \citep{2016A&A...585A.113R}. 
With a goal of understanding the impact of disagreement between the theoretical and actual
telluric transmission spectra on the RV accuracy, we repeated the simulations described in 
Section \ref{s:Doppler} using ``observed" telluric transmission spectra.

\begin{figure}
\includegraphics[width=8.5cm]{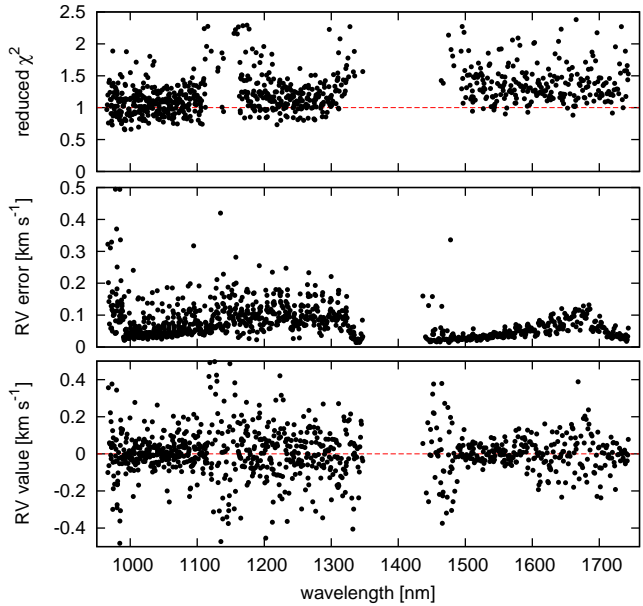}
\caption{
Same as Figure \ref{fig:simu1} except that we used an empirical telluric transmittance in generating
the mock IRD spectra. 
}
\label{fig:simu6}
\end{figure}
Employing the stellar template for the fiducial case, we created 51 mock IRD spectra by Doppler-shifting the template
by $1$ km s$^{-1}$ for each such that the input stellar RV offset ranges from $-25$ km s$^{-1}$ to $+25$ km s$^{-1}$. 
Instead of multiplying the Doppler-shifted templates by the theoretical telluric transmittance (LBLRTM), we multiplied by an empirical telluric
transmittance generated based on the observed spectra of HR 8634 (telluric standard). 
HR 8634 was observed on several different nights between 2018 June and August; 
specifically on UT 2018 August 6, this target was visited 6 times during a night so that 
we can compare telluric spectra taken at different observing conditions (airmass in particular). 
A total of 96 frames were obtained for this target, 
covering the airmass range between $\approx 1.0$ and $\approx 2.0$. 
Since each spectrum of HR 8634 has a limited S/N ratio ($\approx 150-200$ per pixel), 
which may affect the overall quality of the mock spectra, 
we randomly combined multiple ($\approx 5$) frames to generate an empirical telluric spectrum 
for each of the 51 mock data. 
Although this manipulation averages the impact of differing airmass on the telluric lines, 
the combined empirical spectra still exhibit moderate variations in depth and shape due to
different observing conditions. Thus, this numerical experiment also helps us understand 
the impact of random variations in the observing condition on NIR RV measurements.

\begin{figure}
\includegraphics[width=8.5cm]{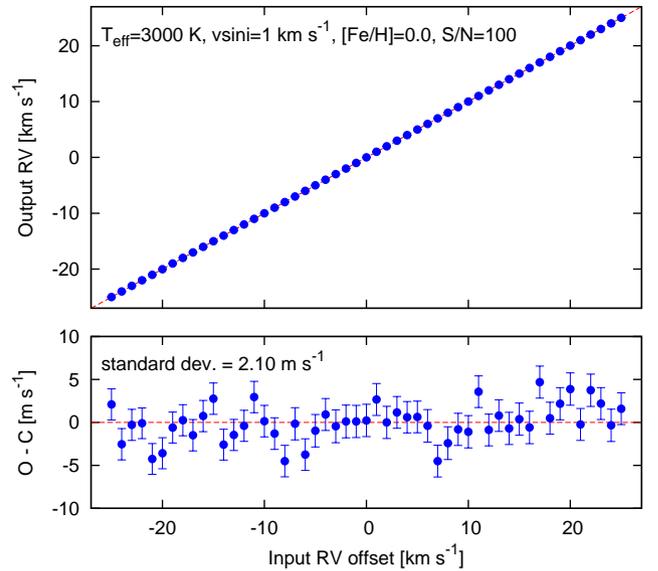}
\caption{
Same as Figure \ref{fig:simu2} except that we used an empirical telluric transmittance in generating
the mock IRD spectra. 
}
\label{fig:simu7}
\end{figure}
Putting these mock IRD spectra into our RV fitting routine, we simulated the RV analysis
for the 51 mock spectra. 
In fitting each mock spectrum, we used the library of ``theoretical" telluric transmittance 
for $T(\mathbf{A};\lambda)$ in Equation (\ref{eq:2}) and optimized the relevant telluric
parameters simultaneously with $v_\star$, 
mimicking the scenario that we analyze real observed spectra. 
Figure \ref{fig:simu6} presents the RV-fit results of individual segments for one 
of the mock spectra ($v_\star=0$ km s$^{-1}$). 
When compared with Figure \ref{fig:simu1}, RV values for individual segments (bottom panel)
exhibit a larger scatter in Figure \ref{fig:simu6}. In many segments that include strong
telluric lines (e.g., $1115\,\mathrm{nm}<\lambda<1165\,\mathrm{nm}$,
$1320\,\mathrm{nm}<\lambda<1495\,\mathrm{nm}$),
the fits did not converge (reduced $\chi^2>3$), or RV values show a very large scatter even 
in case of convergence. As expected, the reduced $\chi^2$ values (top panel) in Figure
\ref{fig:simu6} are generally worse than in Figure \ref{fig:simu1}
due to the disagreement between the theoretical and actual telluric lines. 
Figure \ref{fig:simu6} indicates that the reduced $\chi^2$ value is higher in the $H$-band, 
which is probably related to the fact that for IRD spectra of M dwarfs, S/N ratios are
significantly higher in the $H$-band ($150-200$) than in the $Y$-band ($\approx 100$), 
and the same level of fractional disagreements between the theoretical and observed telluric
absorptions and/or between the model (polynomial) and observed continuum 
leads to a larger $\chi^2$ difference in the $H$-band.

The output RV values as a function of the input RV offset for the 51 mock spectra are plotted
in Figure \ref{fig:simu7}. The bottom panel of the same figure show the residuals between 
the input and output RVs in m s$^{-1}$. The standard deviation of the residuals was found to 
be $2.10$ m s$^{-1}$, which was larger than that in Section \ref{s:Doppler}
by about $28\,\%$. The mean internal error returned by the analysis pipeline was 
$1.85$ m s$^{-1}$, suggesting that the disagreement between the theoretical and actual telluric
lines indeed produces an additional RV scatter of $\approx \sqrt{2.10^2-1.85^2}=0.99$ m s$^{-1}$. 
In order to check if Doppler-shifts of the template relative to the telluric lines lead to 
a systematic variation of resulting RVs, we fitted the $O-C$ residuals in Figure \ref{fig:simu7} 
by a linear function of the input RV offset. 
The best-fit slope was found to be $\kappa=0.053 \pm 0.020$ m s$^{-1}$ (km s$^{-1}$)$^{-1}$. 
Given that the theoretical case was consistent with zero (Section \ref{s:Doppler}), 
this result suggests a hint of correlation between relative telluric-line positions and 
best-fit RV values.   
The application of our pipeline to real data with telluric absorptions could lead to RV
measurements dragged by the positions of telluric lines blended with stellar lines.
Fortunately though, the magnitude of RV shifts by the 
``telluric drag" is only $\approx \pm 1.3$ m s$^{-1}$ even for the worst cases 
(RV offset $=\pm 25$ km s$^{-1}$). In addition, post-processing of the RV data may be able 
to suppress this systematic error in the real data analysis, 
provided that we have a sufficiently large number of RV points. 
For reference, when we subtract the best-fit RV trend (slope) in the residuals from 
the original RVs, the standard deviation becomes $1.94$ m s$^{-1}$, almost consistent 
with the mean internal error.

\section{Tests of the RV Analysis Pipeline: On-sky Performances}\label{s:onsky}

\subsection{Targets and Analyses}

\begin{figure}
\includegraphics[width=8.5cm]{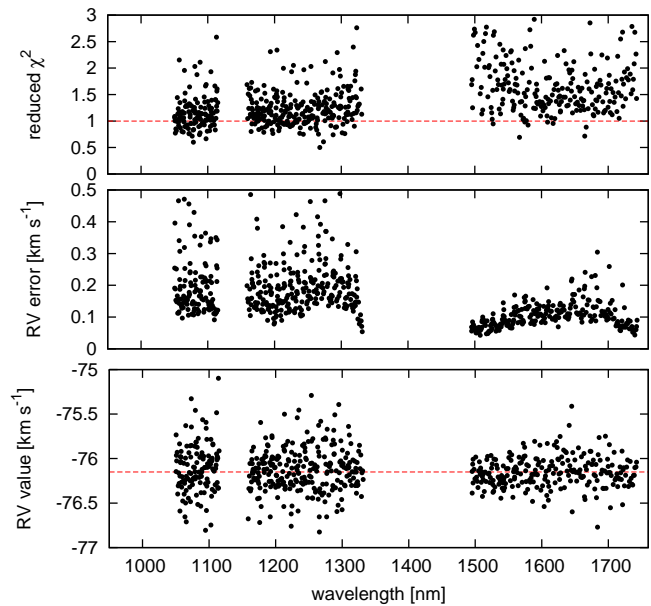}
\caption{
Sample result of the RV analysis for an observed spectrum of TRAPPIST-1. 
RV values (relative to the template), their errors, and reduced $\chi^2$ are plotted for individual segments, 
from bottom to top, respectively. 
}
\label{fig:seg_trappist1}
\end{figure}
As a last set of tests on our RV analysis pipeline, we analyzed the actual spectra 
obtained by IRD. 
The standard stars for RV measurements included GJ 699 (Barnard's star), which is known to host an exoplanet 
\citep{2018Natur.563..365R}, but since its RV variation by barycenter motion of the star due 
to the planet is well determined, we can use the star for IRD's demonstration. 
We also observed TRAPPIST-1 during the open-use programs (proposal ID's: S18B-114, UH-37C, UH-37A) between 2018 August and 2019 July \citep[see][]{2020ApJ...890L..27H}.  
Integration times were set to $45-120$ sec for GJ 699 and $300-1200$ sec for TRAPPIST-1, respectively. 
GJ 699 (Barnard's star) was observed at three epochs (nights) in 2018 June 
and four epochs in 2018 August, while TRAPPIST-1 was observed at four epochs between 2018 
August and 2019 July. 
The extracted 1D spectra had the S/N ratios of $70-180$ and $15-35$ per pixel around 1000 nm for GJ 699 and TRAPPIST-1, respectively.  
As a telluric standard, we observed HR 8634, which is a B8 star having a featureless 
spectrum (rapid rotator) and requiring a very short integration time ($\approx 10$ sec).

Following the procedure outlined in Figure \ref{fig:chart}, we first extracted the instantaneous
IPs for all the spectral segments between $1050\,\mathrm{nm}$ and $1730\,\mathrm{nm}$ for 
each frame, using the LFC spectrum. 
We then deconvolved each stellar spectrum with those IPs, as well as removed the telluric 
absorptions by theoretical-model fitting or dividing by the normalized spectrum of the 
telluric standard star when available. Each deconvolved spectrum was Doppler-shifted to 
the stellar rest frame based on the cross-correlation between a small telluric-free segment 
of the spectrum and a theoretical stellar template (BT-SETTL). 
Checking the barycentric-correction velocity for each frame, we carefully selected a set 
of frames ($>10$ for each target) so that the barycentric motion of Earth leads to the 
largest range of shifts in stellar line positions with respect to the telluric lines, 
whose positions are almost constant in time. 
Those Doppler-shifted frames were eventually median combined to create a high S/N, 
telluric-free stellar template for each target (Figure \ref{fig:templates}) to 
use in the RV analysis (Equation \ref{eq:2}).

\begin{figure*}
\begin{center}
\includegraphics[width=15cm]{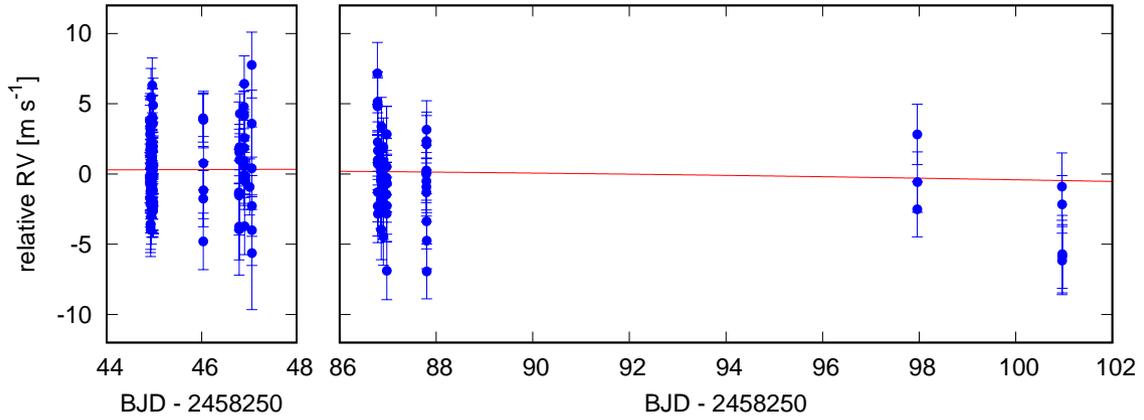}
\end{center}
\caption{
Relative RVs for 176 IRD spectra of GJ 699, analyzed by our pipeline. 
The red solid line represents the Keplerian orbit of GJ 699b \citep{2018Natur.563..365R}. 
}
\label{fig:gj699}
\end{figure*}
Putting each stellar spectrum along with corresponding IPs and the stellar template
into the RV-fit module of the pipeline, we measured the RVs for individual spectral 
segments spanning $\Delta \lambda=0.7-1$ nm. A sample of the RV-fit results (individual 
segments) for TRAPPIST-1 is presented in Figure \ref{fig:seg_trappist1}. 
As predicted in Section \ref{s:telluric}, the reduced $\chi^2$ values are generally worse 
in the $H-$band segments most likely due to the disagreement between the observed telluric 
lines and theoretical telluric spectrum used in the fit. The RV uncertainties returned 
by the pipeline can be compared with the expected
RV precision from numerical simulations for a similar type of star. 
For the specific frame ($\mathrm{ID}=19316$) presented in Figure \ref{fig:seg_trappist1}, 
the internal RV error was $4.98$ m s$^{-1}$
with the S/N ratio being $\approx 21$ per pixel at 1000 nm, in which the corresponding readout noise is taken into account. 
Assuming that the temperature and rotation velocity of TRAPPIST-1 are respectively $T_\mathrm{eff}=2559$ K \citep{2017Natur.542..456G} and $v\sin i \approx 1.5$ km s$^{-1}$\citep{2020ApJ...890L..27H}, 
the expected RV precision according to Figure \ref{fig:simu4}
is $4.8-7.1$ m s$^{-1}$ in case of the solar metallicity, which is fully consistent with the observed one. 
A total of 176 and 105 frames were analyzed for GJ 699 and TRAPPIST-1, respectively. 

\subsection{RV Result: GJ 699}
\begin{figure}
\includegraphics[width=8.5cm]{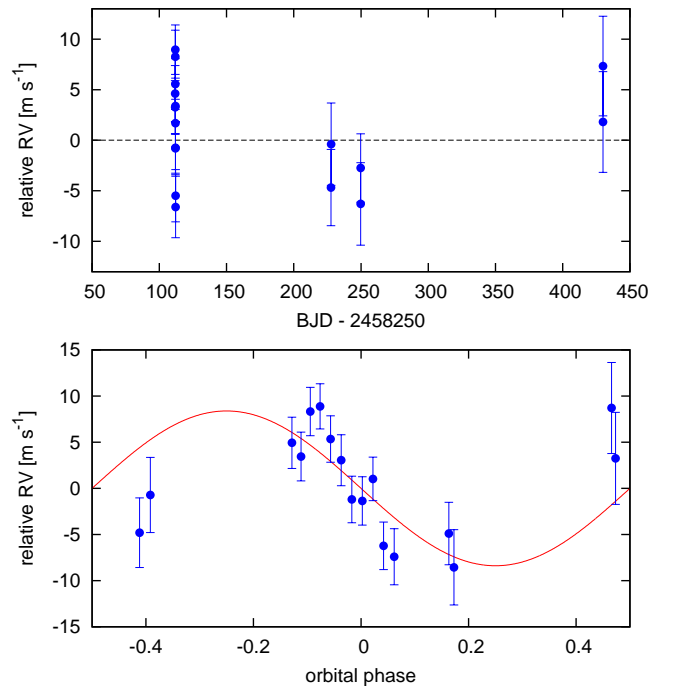}
\caption{
{\it (Upper)} Observed RVs of TRAPPIST-1 (blue points) as a function of BJD$_\mathrm{TDB}$. 
The RV data on $\mathrm{BJD}-2458250\approx 112$ and 228 were binned so that plotted RV 
points have similar RV errors ($3-5$ m s$^{-1}$). 
{\it (Lower)} RVs folded by the orbital period of TRAPPIST-1b ($P=1.51087$ days), 
after subtracting the RV variations
for the other six planets expected from the TTV-based masses \citep{2018A&A...613A..68G}. 
}
\label{fig:trappist1}
\end{figure}
Figure \ref{fig:gj699} plots relative RVs of GJ 699 as a function of time; 
in the plot, the barycentric velocities of Earth were subtracted by using the 
{\tt TEMPO2} software \citep{2006MNRAS.372.1549E}. Given the short integration times,
we adopted the central time of the exposure when we computed the barycentric 
velocity for each frame. 
About half of the data points (82) were obtained on the night of UT 2018 June 25
($\mathrm{BJD}-2458250\approx 45$) to check for the RV stability within a single night. 
The mean internal RV error for the 176 points is 2.06 m s$^{-1}$, while the standard deviation 
of those RVs was found to be 2.72 m s$^{-1}$. 
The Keplerian orbit of GJ 699b, reported by \citet{2018Natur.563..365R}, is drawn 
by the red solid curve in Figure \ref{fig:gj699}.
When we subtract this RV variation by the planet, the standard deviation of
the residual becomes 2.69 m s$^{-1}$. The additional scatter of
$\sqrt{2.69^2-2.06^2}=1.73$ m s$^{-1}$ 
could be ascribed to the effects such as by
(1) instrumental instability (in particular, the ``relative" drift between the stellar and 
LFC spectra), (2) stellar activity, and (3) imperfect removal of telluric lines in the stellar
template (and telluric drag as described in Section \ref{s:telluric}). 
The long-term RV stability of GJ 699 measured by IRD will be discussed in more detail 
in the forthcoming paper (Kotani et al. in prep.). 
Note that the spectra around $\mathrm{BJD}-2458250\approx 101$ were taken at high airmasses
($2.5-2.8$), which might be responsible for a small negative offset seen in 
Figure \ref{fig:gj699}.


\subsection{RV Result: TRAPPIST-1}
For TRAPPIST-1, the integration times were set to only 5 minutes on UT 2018 August 31 
and December 25 in an attempt to observe the Rossiter-McLaughlin effect for this system 
\citep{2020ApJ...890L..27H}, although we were forced to close the dome of the telescope
on December 25 due to very high humidity before the transit started. 
This small integration time yielded much larger errors and scatters in the individual 
RV data points than those of the other nights.
Hence, we binned the RV points on those two nights so that the RV points after binning 
have similar uncertainties ($1\,\mathrm{bin}=7-8$ frames). 
The time scale of transits is only $30-60$ minutes for TRAPPIST-1 planets, 
so the impact of the Rossiter-McLaughlin effect is averaged out by this binning 
(i.e., one binned point approximately covers a full transit). 
The time sequence of TRAPPIST-1's RV result is presented 
in the upper panel of Figure \ref{fig:trappist1}. 
The standard deviation of the plotted RV points is 4.91 m s$^{-1}$, while their mean 
RV error is 3.18 m s$^{-1}$. A part of this additional scatter is definitely ascribed 
to the gravitational perturbations from the seven planets, but the absence of large systematic 
RV variations ($>10$ m s$^{-1}$) suggests that the IRD spectrograph is stable within 
$\approx 5$ m s$^{-1}$ on a time scale up to one year.

For reference, we attempted to fit the orbit of TRAPPIST-1b with the current data set. 
Following \citet{2020ApJ...890L..27H}, we fixed the Keplerian orbits of the other six planets
(c, d, e, f, g, and h) at the ones expected from the TTV-based masses 
\citep{2018A&A...613A..68G}, and fit the RV semi-amplitude $K$ for planet b only, 
assuming a circular orbit. The result of the fit is shown in the lower panel of 
Figure \ref{fig:trappist1}, in which the RV points are folded by the period of TRAPPIST-1b. 
The best-fit RV semi-amplitude was $K=8.4\pm 1.6$ m s$^{-1}$, which was found to be larger 
by $\approx 5$ m s$^{-1}$ than the one expected from TTV 
\citep[$\approx 3$ m s$^{-1}$:][]{2018A&A...613A..68G}\footnote{We confirmed that this result 
is unchanged even if we fit the raw RV points before binning with the inclusion of the 
anomalous RVs due to the Rossiter-McLaughlin effect.}.
The reason for this disagreement is not known, 
but apparently the RV data on UT 2018 August 31 (transit night: orbital phase around zero) 
exhibits a steeper (than expected) slope, as described in \citet{2020ApJ...890L..27H}. 
One reason could be the detector's persistence and/or imperfect removal 
telluric lines in the template, but future observations with IRD or other similar
spectrographs will settle the issue and enable a precise comparison between
the TTV-based masses and RV-based masses. 
We note that one also needs refined ephemerides for all the planets for the RV determination
of accurate planet masses.

\section{Summary and Discussion}\label{s:discussion}

We have described the methodology and pipeline extracting precision RVs from 
NIR high-resolution spectra, and demonstrated the theoretical and observational performances 
of the IRD spectrograph, which simultaneously covers $Y$, $J$, and $H-$bands. 
To account for the characteristics of the IRD spectrographs (e.g., temporal variations of IPs), 
we constructed the RV analysis pipeline with a forward-modeling technique, which 
measures and incorporates the instantaneous variations of telluric lines as well as
segment-by-segment IPs. Our numerical simulations using synthetic spectra (BT-SETTL) 
have shown that for slowly rotating mid-to-late M dwarfs ($v\sin i<2$ km s$^{-1}$), 
which are major targets for the blind Doppler survey in the Subaru Strategic Program
\citep[SSP: e.g.,][]{2018SPIE10702E..11K}, 
IRD can potentially achieve an RV precision of $< 2$ m s$^{-1}$ with a moderate 
S/N ratio ($\gtrsim 100$ per pixel at 1000 nm). 
Through the applications of the new pipeline to the observed spectra, 
we have demonstrated that 
this level of internal precisions is achieved for bright mid-to-late M dwarfs. 
The observed RV variation and scatter for GJ 699 are compatible with 
those reported in the literature \citep{2018Natur.563..365R}, although we 
stress that an additional scatter of $1-2$ m s$^{-1}$ was observed for GJ 699, 
which could be ascribed to the disagreement between observed and theoretical
telluric lines, IRD's instrumental instability, and/or stellar activity.

The methodology presented in this paper is a first-generation approach to extract 
precision RVs from IRD spectra, and we emphasize that there is still plenty room for improvement. 
For instance, in Equation (\ref{eq:2}), we employ theoretical telluric transmission spectra
(LBLRTM) to model each spectrum, but some of the theoretical telluric lines show disagreement
from the observed ones, leading to possible systematic errors in the extracted RVs, 
as explained in Section \ref{s:telluric}. 
A solution to circumvent these systematics is to prepare a set of observed telluric 
spectra by using rapidly rotating (telluric standard) stars and use those empirical telluric spectra 
as a library in modeling each stellar spectrum for RV measurement. This in turns requires 
a large number of observations of telluric standard star(s) to complete various observing
conditions (e.g., airmass, water vapor 
content of the atmosphere, and seasonal variation of the $T-p$ profile on Maunakea). 
A similar approach is explained in \citet{2014SPIE.9149E..05A}, who proposed to 
build a library of absorbances for individual molecular species by the principal 
component analysis.
The library telluric spectra should be free of instrumental broadening, thus requiring the 
deconvolution of IPs for each observed spectrum of the telluric standard star. 
To this end, we have been collecting a number of spectra of telluric standard stars, 
and will continue those observations during the upcoming IRD runs.

One shortcoming of the present technique for RV measurements is that one needs a moderate 
number of spectra ($>10$), each of which should preferentially be well separated 
in terms of observing epochs, to build a high S/N, telluric-free, IP-deconvolved template 
$S(\lambda)$; with a smaller number of observations ($<5$ spectra), one will not be able 
to obtain accurate RVs due to the imperfect (and/or low S/N) template. 
This should also be the case for other RV pipelines using forward-modelings
\citep[e.g.,][]{2018A&A...609A..12Z}, since those techniques essentially require
multiple observations to disentangle stellar lines from the telluric ones.

Fortunately, stars in the main sequence, M dwarfs in particular, are generally characterized 
by a relatively small number of stellar parameters (e.g., absolute magnitude $M_{K_s}$ and 
[Fe/H]) and other relevant parameters such as stellar mass and radius
(surface gravity) can be derived by empirical relations \citep[e.g.,][]{2015ApJ...804...64M}. 
In addition, the SSP blind Doppler survey for mid-to-late M dwarfs is focusing 
exclusively on slowly rotating, magnetically inactive stars to achieve good RV precision and
accuracy \citep{2018SPIE10702E..11K}. 
This means that almost all the stars in the survey sample have similar $v\sin i$, which should 
be less than half of the instrumental resolution of IRD (i.e., $v\sin i\lesssim 2$ km s$^{-1}$). 
In this case, we might be able to substitute the template for RV measurements of a new target 
star (for which only a few spectra are available) with another template generated for a
similar-type star in terms of $T_\mathrm{eff}$ and [Fe/H], having a large number of spectra. 
This attempt to substitute the stellar template is also under progress.

We have focused on RV measurements for spectra obtained by Subaru/IRD, but the techniques
and algorithms described in the present paper can be applied to the NIR spectra 
taken by other Doppler instruments 
such as CARMENES \citep{2016SPIE.9908E..12Q}, HPF \citep{{2014SPIE.9147E..1GM}}, 
and SPIRou \citep{2014SPIE.9147E..15A}. 
Those instruments cover different wavelength regions and use different sources of 
simultaneous wavelength calibration (e.g., a Fabry-Perot based calibration source), 
but we expect that our pipeline can be applied to those instruments with a moderate level 
of tuning in relevant parameters and the telluric library. 
A comparison between the RVs derived by different pipelines would allow us to identify 
possible systematics in RV measurements and gain insight into the origin of
instrumental/telluric/astrophysical correlated noise.


\begin{ack}
We thank all the members of the IRD consortium for helpful discussions on this project. 
The data analysis was carried out, in part, on the Multi-wavelength
Data Analysis System operated by the Astronomy Data Center (ADC),
National Astronomical Observatory of Japan.  This work is supported
by JSPS KAKENHI Grant Numbers 16K17660, 19K14783, 18H05442, 15H02063, 
and 22000005, and by the Astrobiology Center Program of
National Institutes of Natural Sciences (NINS) (Grant Number AB311017). 
We are very grateful to the referee, Mathias Zechmeister, for carefully reading 
our manuscript and providing many insightful comments on our results.  
\end{ack}







\begin{thebibliography}{}
\expandafter\ifx\csname natexlab\endcsname\relax\def\natexlab#1{#1}\fi

\bibitem[{{Allard} {et~al.}(2013){Allard}, {Homeier}, {Freytag},
  {Schaffenberger}, {}, \& {Rajpurohit}}]{2013MSAIS..24..128A}
{Allard}, F., {Homeier}, D., {Freytag}, B., {et~al.} 2013, Memorie della
  Societa Astronomica Italiana Supplementi, 24, 128

\bibitem[{{Anglada-Escud{\'e}} \& {Butler}(2012)}]{2012ApJS..200...15A}
{Anglada-Escud{\'e}}, G., \& {Butler}, R.~P. 2012, \apjs, 200, 15

\bibitem[{{Artigau} {et~al.}(2014{\natexlab{a}}){Artigau}, {Kouach}, {Donati},
  {Doyon}, {Delfosse}, {Baratchart}, {Lacombe}, {Moutou}, {Rabou}, {Par{\`e}s},
  {Micheau}, {Thibault}, {Reshetov}, {Dubois}, {Hernandez}, {Vall{\'e}e},
  {Wang}, {Dolon}, {Pepe}, {Bouchy}, {Striebig}, {H{\'e}nault}, {Loop},
  {Saddlemyer}, {Barrick}, {Vermeulen}, {Dupieux}, {H{\'e}brard}, {Boisse},
  {Martioli}, {Alencar}, {do Nascimento}, \& {Figueira}}]{2014SPIE.9147E..15A}
{Artigau}, {\'E}., {Kouach}, D., {Donati}, J.-F., {et~al.} 2014{\natexlab{a}},
  in \procspie, Vol. 9147, Ground-based and Airborne Instrumentation for
  Astronomy V, 914715

\bibitem[{{Artigau} {et~al.}(2014{\natexlab{b}}){Artigau}, {Astudillo-Defru},
  {Delfosse}, {Bouchy}, {Bonfils}, {Lovis}, {Pepe}, {Moutou}, {Donati},
  {Doyon}, \& {Malo}}]{2014SPIE.9149E..05A}
{Artigau}, {\'E}., {Astudillo-Defru}, N., {Delfosse}, X., {et~al.}
  2014{\natexlab{b}}, in Society of Photo-Optical Instrumentation Engineers
  (SPIE) Conference Series, Vol. 9149, \procspie, 914905

\bibitem[{{Asensio Ramos} \& {Petit}(2015)}]{2015A&A...583A..51A}
{Asensio Ramos}, A., \& {Petit}, P. 2015, \aap, 583, A51

\bibitem[{{Astudillo-Defru} {et~al.}(2017){Astudillo-Defru}, {Forveille},
  {Bonfils}, {S{\'e}gransan}, {Bouchy}, {Delfosse}, {Lovis}, {Mayor}, {Murgas},
  {Pepe}, {Santos}, {Udry}, \& {W{\"u}nsche}}]{2017A&A...602A..88A}
{Astudillo-Defru}, N., {Forveille}, T., {Bonfils}, X., {et~al.} 2017, \aap,
  602, A88

\bibitem[{{Ballard} \& {Johnson}(2016)}]{2016ApJ...816...66B}
{Ballard}, S., \& {Johnson}, J.~A. 2016, \apj, 816, 66

\bibitem[{{Bean} {et~al.}(2010){Bean}, {Seifahrt}, {Hartman}, {Nilsson},
  {Wiedemann}, {Reiners}, {Dreizler}, \& {Henry}}]{2010ApJ...713..410B}
{Bean}, J.~L., {Seifahrt}, A., {Hartman}, H., {et~al.} 2010, \apj, 713, 410

\bibitem[{{Bean} {et~al.}(2006){Bean}, {Sneden}, {Hauschildt}, {Johns-Krull},
  \& {Benedict}}]{2006ApJ...652.1604B}
{Bean}, J.~L., {Sneden}, C., {Hauschildt}, P.~H., {Johns-Krull}, C.~M., \&
  {Benedict}, G.~F. 2006, \apj, 652, 1604

\bibitem[{{Beichman} {et~al.}(2019){Beichman}, {Hirano}, {David}, {Kotani},
  {Hillenbrand}, {Vasisht}, {Ciardi}, {Harakawa}, {Kudo}, {Omiya}, {Kuzuhara},
  \& {Tamura}}]{2019RNAAS...3...89B}
{Beichman}, C., {Hirano}, T., {David}, T.~J., {et~al.} 2019, Research Notes of
  the American Astronomical Society, 3, 89

\bibitem[{{Bonfils} {et~al.}(2013){Bonfils}, {Delfosse}, {Udry}, {Forveille},
  {Mayor}, {Perrier}, {Bouchy}, {Gillon}, {Lovis}, {Pepe}, {Queloz}, {Santos},
  {S{\'e}gransan}, \& {Bertaux}}]{2013A&A...549A.109B}
{Bonfils}, X., {Delfosse}, X., {Udry}, S., {et~al.} 2013, \aap, 549, A109

\bibitem[{{Butler} {et~al.}(1996){Butler}, {Marcy}, {Williams}, {McCarthy},
  {Dosanjh}, \& {Vogt}}]{1996PASP..108..500B}
{Butler}, R.~P., {Marcy}, G.~W., {Williams}, E., {et~al.} 1996, \pasp, 108, 500

\bibitem[{{Clough} {et~al.}(2005){Clough}, {Shephard}, {Mlawer}, {Delamere},
  {Iacono}, {Cady-Pereira}, {Boukabara}, \& {Brown}}]{2005JQSRT..91..233C}
{Clough}, S.~A., {Shephard}, M.~W., {Mlawer}, E.~J., {et~al.} 2005, \jqsrt, 91,
  233

\bibitem[{{Coggins} {et~al.}(1994){Coggins}, {Fullton}, \&
  {Carney}}]{1994rhis.conf...24C}
{Coggins}, J.~M., {Fullton}, L.~K., \& {Carney}, B.~W. 1994, in The Restoration
  of HST Images and Spectra - II, ed. R.~J. {Hanisch} \& R.~L. {White}, 24

\bibitem[{{Donati} {et~al.}(1997){Donati}, {Semel}, {Carter}, {Rees}, \&
  {Collier Cameron}}]{1997MNRAS.291..658D}
{Donati}, J.-F., {Semel}, M., {Carter}, B.~D., {Rees}, D.~E., \& {Collier
  Cameron}, A. 1997, \mnras, 291, 658

\bibitem[{{Dressing} \& {Charbonneau}(2013)}]{2013ApJ...767...95D}
{Dressing}, C.~D., \& {Charbonneau}, D. 2013, \apj, 767, 95

\bibitem[{{Dressing} \& {Charbonneau}(2015)}]{2015ApJ...807...45D}
---. 2015, \apj, 807, 45

\bibitem[{{Dressing} {et~al.}(2017){Dressing}, {Vanderburg}, {Schlieder},
  {Crossfield}, {Knutson}, {Newton}, {Ciardi}, {Fulton}, {Gonzales}, {Howard},
  {Isaacson}, {Livingston}, {Petigura}, {Sinukoff}, {Everett}, {Horch}, \&
  {Howell}}]{2017AJ....154..207D}
{Dressing}, C.~D., {Vanderburg}, A., {Schlieder}, J.~E., {et~al.} 2017, \aj,
  154, 207

\bibitem[{{Edwards} {et~al.}(2006){Edwards}, {Hobbs}, \&
  {Manchester}}]{2006MNRAS.372.1549E}
{Edwards}, R.~T., {Hobbs}, G.~B., \& {Manchester}, R.~N. 2006, \mnras, 372,
  1549

\bibitem[{{Evans} {et~al.}(2015){Evans}, {Aigrain}, {Gibson}, {Barstow},
  {Amundsen}, {Tremblin}, \& {Mourier}}]{2015MNRAS.451..680E}
{Evans}, T.~M., {Aigrain}, S., {Gibson}, N., {et~al.} 2015, \mnras, 451, 680

\bibitem[{{Gaidos} {et~al.}(2016){Gaidos}, {Mann}, {Kraus}, \&
  {Ireland}}]{2016MNRAS.457.2877G}
{Gaidos}, E., {Mann}, A.~W., {Kraus}, A.~L., \& {Ireland}, M. 2016, \mnras,
  457, 2877

\bibitem[{{Gibson} {et~al.}(2012){Gibson}, {Aigrain}, {Roberts}, {Evans},
  {Osborne}, \& {Pont}}]{2012MNRAS.419.2683G}
{Gibson}, N.~P., {Aigrain}, S., {Roberts}, S., {et~al.} 2012, \mnras, 419, 2683

\bibitem[{{Gillon} {et~al.}(2017){Gillon}, {Triaud}, {Demory}, {Jehin}, {Agol},
  {Deck}, {Lederer}, {de Wit}, {Burdanov}, {Ingalls}, {Bolmont}, {Leconte},
  {Raymond}, {Selsis}, {Turbet}, {Barkaoui}, {Burgasser}, {Burleigh}, {Carey},
  {Chaushev}, {Copperwheat}, {Delrez}, {Fernandes}, {Holdsworth}, {Kotze}, {Van
  Grootel}, {Almleaky}, {Benkhaldoun}, {Magain}, \&
  {Queloz}}]{2017Natur.542..456G}
{Gillon}, M., {Triaud}, A.~H.~M.~J., {Demory}, B.-O., {et~al.} 2017, \nat, 542,
  456

\bibitem[{{Gray}(2005)}]{2005oasp.book.....G}
{Gray}, D.~F. 2005, {The Observation and Analysis of Stellar Photospheres}

\bibitem[{{Grimm} {et~al.}(2018){Grimm}, {Demory}, {Gillon}, {Dorn}, {Agol},
  {Burdanov}, {Delrez}, {Sestovic}, {Triaud}, {Turbet}, {Bolmont}, {Caldas},
  {de Wit}, {Jehin}, {Leconte}, {Raymond}, {Van Grootel}, {Burgasser}, {Carey},
  {Fabrycky}, {Heng}, {Hernandez}, {Ingalls}, {Lederer}, {Selsis}, \&
  {Queloz}}]{2018A&A...613A..68G}
{Grimm}, S.~L., {Demory}, B.-O., {Gillon}, M., {et~al.} 2018, \aap, 613, A68

\bibitem[{{Hayano} {et~al.}(2008){Hayano}, {Takami}, {Guyon}, {Oya}, {Hattori},
  {Saito}, {Watanabe}, {Murakami}, {Minowa}, {Ito}, {Colley}, {Eldred},
  {Golota}, {Dinkins}, {Kashikawa}, \& {Iye}}]{2008SPIE.7015E..10H}
{Hayano}, Y., {Takami}, H., {Guyon}, O., {et~al.} 2008, \procspie, Vol. 7015,
  {Current status of the laser guide star adaptive optics system for Subaru
  Telescope}, 701510

\bibitem[{{Hirano} {et~al.}(2015){Hirano}, {Masuda}, {Sato}, {Benomar},
  {Takeda}, {Omiya}, {Harakawa}, \& {Kobayashi}}]{2015ApJ...799....9H}
{Hirano}, T., {Masuda}, K., {Sato}, B., {et~al.} 2015, \apj, 799, 9

\bibitem[{{Hirano} {et~al.}(2011){Hirano}, {Suto}, {Winn}, {Taruya}, {Narita},
  {Albrecht}, \& {Sato}}]{2011ApJ...742...69H}
{Hirano}, T., {Suto}, Y., {Winn}, J.~N., {et~al.} 2011, \apj, 742, 69

\bibitem[{{Hirano} {et~al.}(2018){Hirano}, {Dai}, {Gandolfi}, {Fukui},
  {Livingston}, {Miyakawa}, {Endl}, {Cochran}, {Alonso-Floriano}, {Kuzuhara},
  {Montes}, {Ryu}, {Albrecht}, {Barragan}, {Cabrera}, {Csizmadia}, {Deeg},
  {Eigm{\"u}ller}, {Erikson}, {Fridlund}, {Grziwa}, {Guenther}, {Hatzes},
  {Korth}, {Kudo}, {Kusakabe}, {Narita}, {Nespral}, {Nowak}, {P{\"a}tzold},
  {Palle}, {Persson}, {Prieto-Arranz}, {Rauer}, {Ribas}, {Sato}, {Smith},
  {Tamura}, {Tanaka}, {Van Eylen}, \& {Winn}}]{2018AJ....155..127H}
{Hirano}, T., {Dai}, F., {Gandolfi}, D., {et~al.} 2018, \aj, 155, 127

\bibitem[{{Hirano} {et~al.}(2020){Hirano}, {Gaidos}, {Winn}, {Dai}, {Fukui},
  {Kuzuhara}, {Kotani}, {Tamura}, {Hjorth}, {Albrecht}, {Huber}, {Bolmont},
  {Harakawa}, {Hodapp}, {Ishizuka}, {Jacobson}, {Konishi}, {Kudo}, {Kurokawa},
  {Nishikawa}, {Omiya}, {Serizawa}, {Ueda}, \& {Weiss}}]{2020ApJ...890L..27H}
{Hirano}, T., {Gaidos}, E., {Winn}, J.~N., {et~al.} 2020, \apjl, 890, L27

\bibitem[{{Jones} {et~al.}(2013){Jones}, {Noll}, {Kausch}, {Szyszka}, \&
  {Kimeswenger}}]{2013A&A...560A..91J}
{Jones}, A., {Noll}, S., {Kausch}, W., {Szyszka}, C., \& {Kimeswenger}, S.
  2013, \aap, 560, A91

\bibitem[{{Kashiwagi} {et~al.}(2016){Kashiwagi}, {Kurokawa}, {Okuyama}, {Mori},
  {Tanaka}, {Yamamoto}, \& {Hirano}}]{2016OExpr..24.8120K}
{Kashiwagi}, K., {Kurokawa}, T., {Okuyama}, Y., {et~al.} 2016, Optics Express,
  24, 8120

\bibitem[{{Kerber} {et~al.}(2008){Kerber}, {Nave}, \&
  {Sansonetti}}]{Kerber_2008_ThAr}
{Kerber}, F., {Nave}, G., \& {Sansonetti}, C.~J. 2008, Astrophysical Journal
  Supplement Series, 178, 374

\bibitem[{{Kokubo} {et~al.}(2016){Kokubo}, {Mori}, {Kurokawa}, {Kashiwagi},
  {Tanaka}, {Kotani}, {Nishikawa}, \& {Tamura}}]{2016SPIE.9912E..1RK}
{Kokubo}, T., {Mori}, T., {Kurokawa}, T., {et~al.} 2016, in \procspie, Vol.
  9912, Advances in Optical and Mechanical Technologies for Telescopes and
  Instrumentation II, 99121R

\bibitem[{{Kopparapu} {et~al.}(2016){Kopparapu}, {Wolf}, {Haqq-Misra}, {Yang},
  {Kasting}, {Meadows}, {Terrien}, \& {Mahadevan}}]{2016ApJ...819...84K}
{Kopparapu}, R.~k., {Wolf}, E.~T., {Haqq-Misra}, J., {et~al.} 2016, \apj, 819,
  84

\bibitem[{{Kotani} {et~al.}(2014){Kotani}, {Tamura}, {Suto}, {Nishikawa},
  {Sato}, {Aoki}, {Usuda}, {Kurokawa}, {Kashiwagi}, {Nishiyama}, {Ikeda},
  {Hall}, {Hodapp}, {Hashimoto}, {Morino}, {Okuyama}, {Tanaka}, {Suzuki},
  {Inoue}, {Kwon}, {Suenaga}, {Oh}, {Baba}, {Narita}, {Kokubo}, {Hayano},
  {Izumiura}, {Kambe}, {Kudo}, {Kusakabe}, {Ikoma}, {Hori}, {Omiya}, {Genda},
  {Fukui}, {Fujii}, {Guyon}, {Harakawa}, {Hayashi}, {Hidai}, {Hirano},
  {Kuzuhara}, {Machida}, {Matsuo}, {Nagata}, {Onuki}, {Ogihara}, {Takami},
  {Takato}, {Takahashi}, {Tachinami}, {Terada}, {Kawahara}, \&
  {Yamamuro}}]{2014SPIE.9147E..14K}
{Kotani}, T., {Tamura}, M., {Suto}, H., {et~al.} 2014, in \procspie, Vol. 9147,
  Ground-based and Airborne Instrumentation for Astronomy V, 914714

\bibitem[{{Kotani} {et~al.}(2018){Kotani}, {Tamura}, {Nishikawa}, {Ueda},
  {Kuzuhara}, {Omiya}, {Hashimoto}, {Ishizuka}, {Hirano}, {Suto}, {Kurokawa},
  {Kokubo}, {Mori}, {Tanaka}, {Kashiwagi}, {Konishi}, {Kudo}, {Sato},
  {Jacobson}, {Hodapp}, {Hall}, {Aoki}, {Usuda}, {Nishiyama}, {Nakajima},
  {Ikeda}, {Yamamuro}, {Morino}, {Baba}, {Hosokawa}, {Ishikawa}, {Narita},
  {Kokubo}, {Hayano}, {Izumiura}, {Kambe}, {Kusakabe}, {Kwon}, {Ikoma}, {Hori},
  {Genda}, {Fukui}, {Fujii}, {Kawahara}, {Olivier}, {Jovanovic}, {Harakawa},
  {Hayashi}, {Hidai}, {Machida}, {Matsuo}, {Nagata}, {Ogihara}, {Takami},
  {Takato}, {Terada}, \& {Oh}}]{2018SPIE10702E..11K}
{Kotani}, T., {Tamura}, M., {Nishikawa}, J., {et~al.} 2018, in \procspie, Vol.
  10702, Ground-based and Airborne Instrumentation for Astronomy VII, 1070211

\bibitem[{{Kuzuhara} {et~al.}(2018){Kuzuhara}, {Hirano}, {Kotani}, {Ishizuka},
  {Omiya}, {Konishi}, {Kudo}, {Nishikawa}, {Ueda}, {Hosokawa}, {Kusakabe},
  {Kurokawa}, {Kokubo}, {Mori}, {Tanaka}, {Jacobson}, {Hodapp}, \&
  {Tamura}}]{2018SPIE10702E..60K}
{Kuzuhara}, M., {Hirano}, T., {Kotani}, T., {et~al.} 2018, in \procspie, Vol.
  10702, Ground-based and Airborne Instrumentation for Astronomy VII, 1070260

\bibitem[{{Mahadevan} {et~al.}(2014){Mahadevan}, {Ramsey}, {Terrien},
  {Halverson}, {Roy}, {Hearty}, {Levi}, {Stefansson}, {Robertson}, {Bender},
  {Schwab}, \& {Nelson}}]{2014SPIE.9147E..1GM}
{Mahadevan}, S., {Ramsey}, L.~W., {Terrien}, R., {et~al.} 2014, in \procspie,
  Vol. 9147, Ground-based and Airborne Instrumentation for Astronomy V, 91471G

\bibitem[{{Mann} {et~al.}(2015){Mann}, {Feiden}, {Gaidos}, {Boyajian}, \& {von
  Braun}}]{2015ApJ...804...64M}
{Mann}, A.~W., {Feiden}, G.~A., {Gaidos}, E., {Boyajian}, T., \& {von Braun},
  K. 2015, \apj, 804, 64

\bibitem[{{Noll} {et~al.}(2012){Noll}, {Kausch}, {Barden}, {Jones}, {Szyszka},
  {Kimeswenger}, \& {Vinther}}]{2012A&A...543A..92N}
{Noll}, S., {Kausch}, W., {Barden}, M., {et~al.} 2012, \aap, 543, A92

\bibitem[{{Pepe} {et~al.}(2002){Pepe}, {Mayor}, {Galland}, {Naef}, {Queloz},
  {Santos}, {Udry}, \& {Burnet}}]{2002A&A...388..632P}
{Pepe}, F., {Mayor}, M., {Galland}, F., {et~al.} 2002, \aap, 388, 632

\bibitem[{{Press} {et~al.}(2002){Press}, {Teukolsky}, {Vetterling}, \&
  {Flannery}}]{2002nrc..book.....P}
{Press}, W.~H., {Teukolsky}, S.~A., {Vetterling}, W.~T., \& {Flannery}, B.~P.
  2002, {Numerical recipes in C++ : the art of scientific computing}

\bibitem[{{Quirrenbach} {et~al.}(2016){Quirrenbach}, {Amado}, {Caballero},
  {Mundt}, {Reiners}, {Ribas}, {Seifert}, {Abril}, {Aceituno},
  {Alonso-Floriano}, {Anwand-Heerwart}, {Azzaro}, {Bauer}, {Barrado},
  {Becerril}, {Bejar}, {Benitez}, {Berdinas}, {Brinkm{\"o}ller}, {Cardenas},
  {Casal}, {Claret}, {Colom{\'e}}, {Cortes-Contreras}, {Czesla}, {Doellinger},
  {Dreizler}, {Feiz}, {Fernandez}, {Ferro}, {Fuhrmeister}, {Galadi},
  {Gallardo}, {G{\'a}lvez-Ortiz}, {Garcia-Piquer}, {Garrido}, {Gesa},
  {G{\'o}mez Galera}, {Gonz{\'a}lez Hern{\'a}ndez}, {Gonzalez Peinado},
  {Gr{\"o}zinger}, {Gu{\`a}rdia}, {Guenther}, {de Guindos}, {Hagen}, {Hatzes},
  {Hauschildt}, {Helmling}, {Henning}, {Hermann}, {Hern{\'a}ndez Arabi},
  {Hern{\'a}ndez Casta{\~n}o}, {Hern{\'a}ndez Hernando}, {Herrero}, {Huber},
  {Huber}, {Huke}, {Jeffers}, {de Juan}, {Kaminski}, {Kehr}, {Kim}, {Klein},
  {Kl{\"u}ter}, {K{\"u}rster}, {Lafarga}, {Lara}, {Lamert}, {Laun},
  {Launhardt}, {Lemke}, {Lenzen}, {Llamas}, {Lopez del Fresno},
  {L{\'o}pez-Puertas}, {L{\'o}pez-Santiago}, {Lopez Salas}, {Magan
  Madinabeitia}, {Mall}, {Mandel}, {Mancini}, {Marin Molina}, {Maroto
  Fern{\'a}ndez}, {Mart{\'{\i}}n}, {Mart{\'{\i}}n-Ruiz}, {Marvin}, {Mathar},
  {Mirabet}, {Montes}, {Morales}, {Morales Mu{\~n}oz}, {Nagel}, {Naranjo},
  {Nowak}, {Palle}, {Panduro}, {Passegger}, {Pavlov}, {Pedraz}, {Perez},
  {P{\'e}rez-Medialdea}, {Perger}, {Pluto}, {Ram{\'o}n}, {Rebolo}, {Redondo},
  {Reffert}, {Reinhart}, {Rhode}, {Rix}, {Rodler}, {Rodr{\'{\i}}guez},
  {Rodr{\'{\i}}guez L{\'o}pez}, {Rohloff}, {Rosich}, {Sanchez Carrasco},
  {Sanz-Forcada}, {Sarkis}, {Sarmiento}, {Sch{\"a}fer}, {Schiller}, {Schmidt},
  {Schmitt}, {Sch{\"o}fer}, {Schweitzer}, {Shulyak}, {Solano}, {Stahl},
  {Storz}, {Tabernero}, {Tala}, {Tal-Or}, {Ulbrich}, {Veredas}, {Vico Linares},
  {Vilardell}, {Wagner}, {Winkler}, {Zapatero Osorio}, {Zechmeister},
  {Ammler-von Eiff}, {Anglada-Escud{\'e}}, {del Burgo}, {Garcia-Vargas},
  {Klutsch}, {Lizon}, {Lopez-Morales}, {Ofir}, {P{\'e}rez-Calpena}, {Perryman},
  {S{\'a}nchez-Blanco}, {Strachan}, {St{\"u}rmer}, {Su{\'a}rez}, {Trifonov},
  {Tulloch}, \& {Xu}}]{2016SPIE.9908E..12Q}
{Quirrenbach}, A., {Amado}, P.~J., {Caballero}, J.~A., {et~al.} 2016, in
  \procspie, Vol. 9908, Ground-based and Airborne Instrumentation for Astronomy
  VI, 990812

\bibitem[{{Reiners} {et~al.}(2010){Reiners}, {Bean}, {Huber}, {Dreizler},
  {Seifahrt}, \& {Czesla}}]{2010ApJ...710..432R}
{Reiners}, A., {Bean}, J.~L., {Huber}, K.~F., {et~al.} 2010, \apj, 710, 432

\bibitem[{{Ribas} {et~al.}(2018){Ribas}, {Tuomi}, {Reiners}, {Butler},
  {Morales}, {Perger}, {Dreizler}, {Rodr{\'\i}guez-L{\'o}pez}, {Gonz{\'a}lez
  Hern{\'a}ndez}, {Rosich}, {Feng}, {Trifonov}, {Vogt}, {Caballero}, {Hatzes},
  {Herrero}, {Jeffers}, {Lafarga}, {Murgas}, {Nelson}, {Rodr{\'\i}guez},
  {Strachan}, {Tal-Or}, {Teske}, {Toledo-Padr{\'o}n}, {Zechmeister},
  {Quirrenbach}, {Amado}, {Azzaro}, {B{\'e}jar}, {Barnes}, {Berdi{\~n}as},
  {Burt}, {Coleman}, {Cort{\'e}s-Contreras}, {Crane}, {Engle}, {Guinan},
  {Haswell}, {Henning}, {Holden}, {Jenkins}, {Jones}, {Kaminski}, {Kiraga},
  {K{\"u}rster}, {Lee}, {L{\'o}pez-Gonz{\'a}lez}, {Montes}, {Morin}, {Ofir},
  {Pall{\'e}}, {Rebolo}, {Reffert}, {Schweitzer}, {Seifert}, {Shectman},
  {Staab}, {Street}, {Su{\'a}rez Mascare{\~n}o}, {Tsapras}, {Wang}, \&
  {Anglada-Escud{\'e}}}]{2018Natur.563..365R}
{Ribas}, I., {Tuomi}, M., {Reiners}, A., {et~al.} 2018, \nat, 563, 365

\bibitem[{{Rodler} {et~al.}(2011){Rodler}, {Del Burgo}, {Witte}, {Helling},
  {Hauschildt}, {Mart{\'\i}n}, {{\'A}lvarez}, \&
  {Deshpande}}]{2011A&A...532A..31R}
{Rodler}, F., {Del Burgo}, C., {Witte}, S., {et~al.} 2011, \aap, 532, A31

\bibitem[{{Rudolf} {et~al.}(2016){Rudolf}, {G{\"u}nther}, {Schneider}, \&
  {Schmitt}}]{2016A&A...585A.113R}
{Rudolf}, N., {G{\"u}nther}, H.~M., {Schneider}, P.~C., \& {Schmitt},
  J.~H.~M.~M. 2016, \aap, 585, A113

\bibitem[{{Sato} {et~al.}(2002){Sato}, {Kambe}, {Takeda}, {Izumiura}, \&
  {Ando}}]{2002PASJ...54..873S}
{Sato}, B., {Kambe}, E., {Takeda}, Y., {Izumiura}, H., \& {Ando}, H. 2002,
  \pasj, 54, 873

\bibitem[{{Tamura} {et~al.}(2012){Tamura}, {Suto}, {Nishikawa}, {Kotani},
  {Sato}, {Aoki}, {Usuda}, {Kurokawa}, {Kashiwagi}, {Nishiyama}, {Ikeda},
  {Hall}, {Hodapp}, {Hashimoto}, {Morino}, {Inoue}, {Mizuno}, {Washizaki},
  {Tanaka}, {Suzuki}, {Kwon}, {Suenaga}, {Oh}, {Narita}, {Kokubo}, {Hayano},
  {Izumiura}, {Kambe}, {Kudo}, {Kusakabe}, {Ikoma}, {Hori}, {Omiya}, {Genda},
  {Fukui}, {Fujii}, {Guyon}, {Harakawa}, {Hayashi}, {Hidai}, {Hirano},
  {Kuzuhara}, {Machida}, {Matsuo}, {Nagata}, {Ohnuki}, {Ogihara}, {Oshino},
  {Suzuki}, {Takami}, {Takato}, {Takahashi}, {Tachinami}, \&
  {Terada}}]{2012SPIE.8446E..1TT}
{Tamura}, M., {Suto}, H., {Nishikawa}, J., {et~al.} 2012, in \procspie, Vol.
  8446, Ground-based and Airborne Instrumentation for Astronomy IV, 84461T

\bibitem[{{Tennyson} \& {Yurchenko}(2018)}]{2018Atoms...6...26T}
{Tennyson}, J., \& {Yurchenko}, S. 2018, Atoms, 6, 26

\bibitem[{{Valenti} \& {Fischer}(2005)}]{2005ApJS..159..141V}
{Valenti}, J.~A., \& {Fischer}, D.~A. 2005, \apjs, 159, 141

\bibitem[{{Valenti} {et~al.}(1998){Valenti}, {Piskunov}, \&
  {Johns-Krull}}]{1998ApJ...498..851V}
{Valenti}, J.~A., {Piskunov}, N., \& {Johns-Krull}, C.~M. 1998, \apj, 498, 851

\bibitem[{{Zechmeister} {et~al.}(2018){Zechmeister}, {Reiners}, {Amado},
  {Azzaro}, {Bauer}, {B{\'e}jar}, {Caballero}, {Guenther}, {Hagen}, {Jeffers},
  {Kaminski}, {K{\"u}rster}, {Launhardt}, {Montes}, {Morales}, {Quirrenbach},
  {Reffert}, {Ribas}, {Seifert}, {Tal-Or}, \& {Wolthoff}}]{2018A&A...609A..12Z}
{Zechmeister}, M., {Reiners}, A., {Amado}, P.~J., {et~al.} 2018, \aap, 609, A12

\end{thebibliography}

\end{document}